\documentclass[preprints,article,accept,moreauthors,pdftex]{Definitions/mdpi} 
\usepackage{amsmath,amsfonts,amssymb,bm}
\usepackage{subfigure}
\usepackage{graphicx}
\usepackage{hyperref}

\newcommand{\bea}{\begin{eqnarray}}
\newcommand{\eea}{\end{eqnarray}}
\newcommand{\be}{\begin{equation}}
\newcommand{\ee}{\end{equation}}

\newcommand{\mbf}{\mathbf}

\firstpage{1} 
\pubvolume{1}
\issuenum{1}
\articlenumber{0}
\pubyear{2021}
\copyrightyear{2020}
\datereceived{} 
\dateaccepted{} 
\datepublished{} 
\hreflink{https://doi.org/} % If needed use \linebreak
\Title{Thermo-magneto-electric transport through a torsion dislocation in a type I Weyl Semimetal}

\TitleCitation{Thermo-magneto-electric transport through a torsion dislocation in a type I Weyl Semimetal}

\Author{Daniel Bonilla $^{1}$, Enrique Mu\~noz $^{1,2,*}$\orcidA{} and Rodrigo Soto-Garrido $^{1}$ \orcidB{}}

\AuthorNames{Daniel Bonilla, Enrique Mu\~noz and Rodrigo Soto-Garrido}

\AuthorCitation{Bonilla, D.; Mu\~noz, E.; Soto-Garrido, R.}
\address{%
$^{1}$ \quad Physics Institute, Pontificia Universidad Cat\'olica de Chile, Avenida Vicu\~na Mackenna 4860, Santiago, Chile.\\
$^{2}$ \quad Research Center for Nanotechnology and Advanced Materials, CIEN-UC, Pontificia Universidad Cat\'olica de Chile, Avenida Vicu\~na Mackenna 4860, Santiago, Chile.
}
\corres{Correspondence: munozt@fis.puc.cl; Tel.:  +56-2-2354-7625 }

\abstract{We study electronic and thermoelectric transport in a type I Weyl semimetal nanojunction, with a torsional dislocation defect, in the presence of an external magnetic field parallel to the dislocation axis. The defect is modeled in a cylindrical geometry, as a combination of a gauge field accounting for torsional strain, and a delta-potential barrier for the lattice mismatch effect. In the Landauer formalism, we find that due to the combination of strain and magnetic field, the electric current exhibits chiral valley-polarization, and the conductance displays the signature of Landau levels. We also compute the thermal transport coefficients, where a high thermopower and a large figure of merit are predicted for the junction.}


\begin{document}
%%%%%%%%%%%%%%%%%%%%%%%%%%%%%%%%%%%%%%%%%%

\section{Introduction}

Since the experimental discovery of topological insulators, there has been an increasing interest in the search for other materials that may exhibit non-trivial topological properties \cite{hasan2010,qi2011,topol_band_theory,vanderbilt,moore}. A remarkable example of three-dimensional gapless topological materials are Weyl semimetals (WSMs). First proposed theoretically, \cite{Xiangang2011,Fang2012,Ruan2016,Dirac,Felser,3dWeyl,Burkov} WSMs were recently discovered experimentally on TaAs crystals\cite{Xu613} and observed in photonic crystals \cite{Lu622}.
In a WSM, the conduction and valence bands touch each other in an even number of points with linear dispersion, referred as Weyl nodes. These nodes are protected from being gapped because they are monopolar sources of Berry curvature, and hence their charge (chirality) is a topological invariant \cite{Burkov}. In the vicinity of these nodes, low energy conducting states can be described as Weyl fermions, i.e. massless quasi-particles with pseudo-relativistic Dirac linear dispersion \cite{Dirac,Felser,3dWeyl,3dWeyl,Burkov}. In addition to their intrinsic electronic spin, in Weyl fermions chirality determines the projection of the spin over their momentum direction, a condition often referred to as ``spin-momentum locked states''. While Type I WSMs fully respect Lorentz covariance, such condition is not satisfied in Type II WSMs, where the Dirac cones are strongly tilted \cite{vanderbilt}.\\

The presence of Weyl nodes in the bulk spectrum determines the emergence of Fermi arcs\cite{Xu613}, the chiral anomaly, and the chiral magnetic effect, among other remarkable properties \cite{vanderbilt}.  Perhaps the most studied is the chiral anomaly, which is the non-conservation of the independent chiral currents in the presence of non-orthogonal electric and magnetic fields. Therefore, considerable attention has been paid to understand the electronic transport properties of WSMs \cite{Hosur2013,Hu2019,Nagaosa2020}. For instance, there are recent works on charge transport \cite{Hosur2012} in the presence of spin-orbit coupled impurities \cite{Liu2017}, electrochemical\cite{Flores2021} and nonlinear transport induced by Berry curvature dipoles \cite{Zeng2021}. Regarding thermoelectric transport in WSMs, it is known that the linear Dirac-type dispersion induces a non-trivial dependence on the chemical potential \cite{Lundgren2014}. Somewhat less explored are the effects of mechanical strain and deformations in WSMs. From the theory perspective, it has been proposed that different sorts of elastic strains can be modeled as gauge fields in WSMs \cite{Cortijo_2015,Cortijo_2016,Arjona_Vozmediano_PRB2018}, similar to the case of graphene. In previous works, we have studied the effects of strain and magnetic field on the electronic \cite{Soto_Garrido_2018,Soto_Garrido_2020} and the thermoelectric \cite{Munoz2019} transport properties of WSMs, using the Landauer ballistic formalism in combination with the quantum mechanical scattering cross-sections\cite{Munoz_2017}. The study of thermoelectric transport properties is a field of permanent interest, not only regarding WSMs but in a wide range of materials. For instance, there is recent literature involving the experimental determination of the thermoelectric properties (in particular the figure of merit ZT) of Cu-Sn–based thiospinel compounds \cite{Bourges_2020}, and SnTe-based materials \cite{Muchtar_2021}.  \\

\begin{figure}[hbt]
\centering
 \includegraphics[width=0.6 \columnwidth]{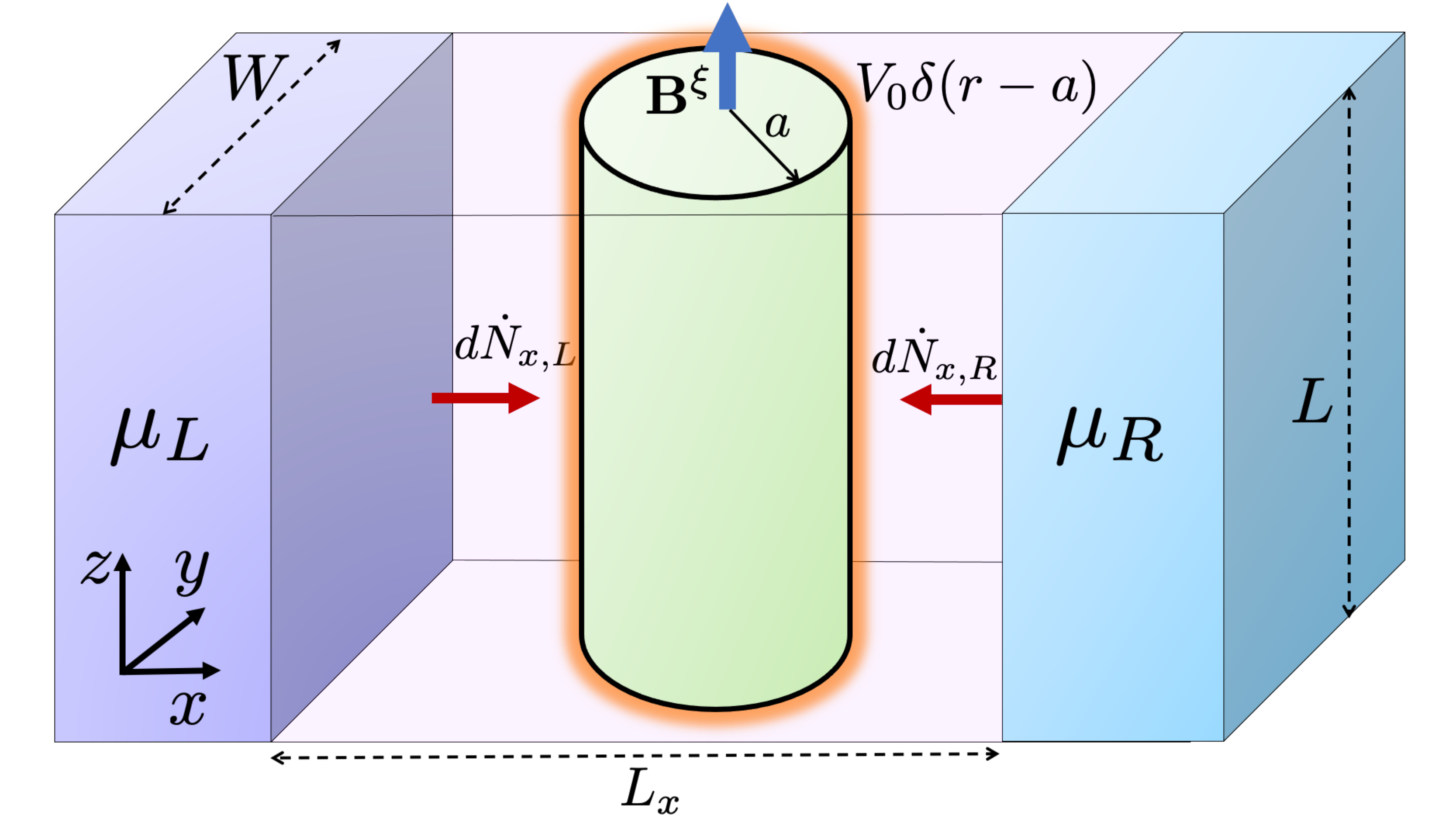}
 \caption{A pictorial description of the system under consideration: A WSM slab of dimensions $L\times W$, with a cylindrical region of radius $a$ submitted to a combination of torsional strain and an external magnetic field $\mbf{B}^{\xi}=(B + \xi B_S)\mbf{\hat{z}}$ and an RDSP on the boundary surface of the cylinder.}
 \label{fig:fig1}
 \end{figure}

This work focuses on the effect of a Repulsive Delta-Shell potential (RDSP), in addition to the torsional strain and the external magnetic field studied early on in Refs. \cite{Munoz2019,Soto_Garrido_2018}, on the thermoelectric transport properties of type I WSMs. The RDSP is a toy model for the surface repulsion produced by the mismatch between the lattices of the strained and the non-strained WSMs. The effect of the delta potential in the context of the Dirac equation is to produce a chiral rotation between the spinors on either side of the boundary that represents the support of the delta function \cite{calkin,Benguria_2000}. The rotation angle is proportional to the strength of the delta barrier and depends on the chirality of the fermion scattered. This RDSP model for the lattice mismatch of the dislocation is combined with a gauge field representation of the torsional strain in a cylindrical geometry. In addition, an external magnetic field directed along the axis of the dislocation is imposed at the junction, as depicted in Fig.~\ref{fig:fig1}.

The paper is organized as follows. In Sec.~Theory we establish the Hamiltonian for the model, and describe each of its contributions. Then, we proceed with the Landauer formulation for transport accross the junction, first analyzing the sole effect of the RDSP that describes the lattice mismatch, and finally for the full system that includes the torsional strain and the external magnetic field at the WSM junction, with mathematical details presented in the Supplementary Material's file. The analysis and discussion of the results are presented in Sec.~Results, with a final summary and conclusions presented in Sec.~Discussion.

%%%%%%%%%%%%%%%%%%%%%%%%%%%%%%%%%%%%%%%%%%
\section{Theory}

\label{sec:Theory}
As a minimal model for a WSM, we start by considering a free Hamiltonian describing Weyl quasiparticles in the vicinity of each of the nodal points with opposite chirality $\xi=\pm 1$,
\bea
H_{\xi}(\mathbf{k}) = v_F \left( \sigma_1 k_x + \sigma_2 k_y  + \xi \sigma_3 k_z\right)
\eea
with $\sigma_j$ ($j = 1,2,3$) the Pauli matrices.
The spectrum of this ``free'' WSM Hamiltonian is given by (for $\lambda = \pm$ the band index)
\bea
E_{\lambda,k}=\lambda\hbar  v_{F}|\mbf{k}|.
\label{eq:free_spectrum}
\eea

As depicted in Fig.~\ref{fig:fig1}, we consider a nanojunction where the WSM is submitted to torsional strain in a cylindrical region of radius $a$, and we further assume that the axial length $L$ satisfies $L\gg a$. As discussed in Ref.~\cite{Arjona_Vozmediano_PRB2018}, the mechanical strain effect can be incorporated as a gauge field $\mathbf{A}_S = B_S/2\left(-y\hat{e}_1 + x\hat{e}_2  \right)$, where the constant $B_S$ plays the role of a pseudo-magnetic field. Moreover, if a true magnetic field is imposed upon the junction along the axis of the dislocation, i.e. $\mathbf{B} = \hat{e}_3 B_0$, then the combination is described by a node-dependent gauge field $\mathbf{A}_{\xi} = B_{\xi}/2\left( -y\hat{e}_1 + x\hat{e}_2 \right)$, with $B_{\xi} = \left(B_0 + \xi B_S\right)$ an effective pseudo-magnetic field. In addition to this combined effect, already discussed in our previous work\cite{Soto_Garrido_2018,Munoz2019}, we here consider also the lattice mismatch near the boundary of the dislocation. As a simple model for this effect, we include a RDSP potential of the form $V_{RD}(r) = V_0\delta(r-a)$. Therefore, the quasi-particle states inside the dislocation region correspond to the solutions of the eigenvalue problem
\bea
\left[H_{\xi}\left(\mathbf{k} + \mathbf{A}_{\xi}  \right) + V_{RD}(r)\right]|\Psi_{n,m}^{(\lambda,\xi)} \rangle
= E_{\lambda,n}^{\xi}| \Psi_{n,m}^{(\lambda,\xi)}.  \rangle
\eea
The spectrum inside the cylindrical region\cite{Munoz_2017} corresponds to relativistic Landau levels with an effective magnetic field $B_{\xi}$ that is node-dependent
\begin{eqnarray}
	E_{\lambda,n}^{\xi} = \lambda \hbar v_F \sqrt{2 n |B_{\xi}|/\tilde{\phi}_0  + k_z^2}, \label{eq:spectrum}
\end{eqnarray}
with $\tilde{\phi}_0=(v_{F}/c)\hbar/e$ a modified magnetic flux quantum expressed in terms of the carrier velocity $v_{F}$.
The effect of the RDSP potential (see Sec. 1 of the Supplementary Material for mathematical details) is to introduce a rotation in the pseudo-spinor components across the dislocation boundary $r=a$, with an ``angle'' $\alpha = V_0/(\hbar v_F)$
\bea
\left.\Psi_{n,m}^{(\lambda,\xi)}(\mathbf{r})\right|_{r\rightarrow a^{+}}=\begin{pmatrix}
 \cos \alpha & -\sin \alpha \\
 \sin \alpha & \cos \alpha
 \end{pmatrix} \left.\Psi_{n,m}^{(\lambda,\xi)}(\mathbf{r})\right|_{r\rightarrow a^{-}}.
 \eea

\subsection{Transmission and Landauer conductance}
In the Landauer formalism, we define an energy-dependent transmission coefficient along the $x$-direction based on the scattering differential cross-section of the junction,
\bea
\bar{T}(E) = \int_{-\pi/2}^{\pi/2}d\phi\cos\phi
\frac{1}{\sigma(E)}\frac{d\sigma}{d\phi},
\eea
where $\sigma(E)$ is the total scattering cross-section at energy $E$.
In what follows, we shall assume that the cylindrical dislocation satisfies $L \gg 1/k_F$. For instance\cite{Neupane2014}, in TaAs where $b\sim 0.08$ $\mathring{\text{A}}^{-1}$ and $v_F\sim 1.3\times 10^{5}$ m/s, we have $1/k_F\sim 9$ $\mathring{\text{A}}$, so even a slab of a few microns is already in the range of validity of this assumption. Moreover, for Cd$_3$As$_2$, $b\sim 0.2$ $\mathring{\text{A}}^{-1}$ and $v_F\sim 1.5\times 10^{6}$ m/s, $1/k_F\sim 0.8$ $\mathring{\text{A}}$ \cite{Neupane2014}, and hence the applicability of this criteria is even more striking in this second example.
Therefore, for $L \gg 1/k_F$, the differential cross-section is given in terms of the scattering phase-shift $\delta_{m}$ for each angular momentum channel $m$\cite{Munoz_2017,Soto_Garrido_2018}, and integrating over the scattering angle (see Sec. 2 of the Supplementary Material for mathematical details) we obtain the corresponding total cross-section\cite{Munoz_2017,Soto_Garrido_2018} $\sigma/L =\frac{4 }{k_{\perp}}\sum_{m=-\infty}^{\infty}\sin^2\delta_{m}$.

Let us first consider the effect of the RDSP only. For this case, the current is expressed in terms of the transmission function $\mathcal{T}(E)$, evaluated at the free energy eigenvalues $E_{\lambda,k_{\perp}}$ defined in Eq.~(\ref{eq:free_spectrum})
\bea
I=2 ev_{F}\sum_{\lambda}\int_{0}^{\infty}dk_{\perp} \mathcal{T}(E_{\lambda,k_{\perp}}) \left[ f_L(E_{\lambda,k_{\perp}}) - f_R(E_{\lambda,k_{\perp}}) \right], \label{eq:node_elec_current_transm_delta}
\eea
where $f_{L/R}(E) = \left(\exp[(E-\mu_{L/R})/(k_B T_{L/R})]+1\right)^{-1}$ are the Fermi-Dirac distributions at the chemical potential $\mu_{L/R}$ and temperature $T_{L/R}$ of the left (L) and right (R) metallic contacts, respectively (see Sec. 3 of the Supplementary Material for mathematical details). The factor of 2 accounts for the (symmetric) contribution from each chiral node $\xi=\pm$ (see Fig.~\ref{fig:tan_phase_shift_delta}). The corresponding expression for the differential conductance $G(T,V) = \left.\partial I/\partial V\right|_T$ through the junction is
\begin{equation}
G(T,V) = 2\frac{e^2 v_F}{k_B T} \sum_{\lambda}\int_{0}^{\infty}dk_{\perp} \mathcal{T}(E_{\lambda,k_{\perp}}) f_L(E_{\lambda,k_{\perp}}) \left[ 1 - f_L(E_{\lambda,k_{\perp}}) \right]. \label{eq:conductance_delta}
\end{equation}

Let us now consider the transmission through the junction in its full level of complexity, i.e. including the RDSP for the lattice mismatch, as well as the torsional strain (included via the gauge field model) and the external magnetic field along the axis of the cylindrical dislocation. For this case, scattering is no longer symmetric for each chirality, as seen in the Landau level spectrum $E_{\lambda,n}^{\xi}$ defined in Eq.~(\ref{eq:spectrum}) and in the corresponding scattering phase shift (Fig.~\ref{fig:tan_phase_shift_delta}). Therefore,
the current for each chirality $\xi=\pm$ is expressed by the transmission function $\mathcal{T}(E)$,
\bea
I_{\xi} = e v_F \sum_{n,\lambda}  \mathcal{T}(E_{\lambda,n}^{\xi})\left[ f_L(E_{\lambda,n}^{\xi}) - f_R(E_{\lambda,n}^{\xi})\right],
\label{eq:node_elec_current_transm_mag_torsion_RDSP}
\eea
with the total current defined by the superposition of both chiral contributions $I = I_{+} + I_{-}$. As before, the differential conductance through the junction is obtained as the voltage-derivative of the expression above, 
\bea
G(T,V) = \frac{e^2 v_F}{k_B T} \sum_{\lambda,n,\xi}\mathcal{T}(E_{\lambda,n}^{\xi})f_L(E_{\lambda,n}^{\xi})\left[
1 - f_L(E_{\lambda,n}^{\xi})\right]. \label{eq:conductance_mag_torsion_RDSP}
\eea

\subsection{Thermoelectric transport coefficients}
\label{subsec:thermo}
The energy current accross the junction arising from each chiral node contribution $\xi=\pm$ is also expressed in terms of the transmission function $\mathcal{T}(E)$ as follows\cite{Munoz2019}
\bea
\dot{U}_{\xi} =  v_F \sum_{n,\lambda} E_{\lambda,n}^{\xi} \mathcal{T}(E_{\lambda,n}^{\xi})\left[ f_L(E_{\lambda,n}^{\xi}) - f_R(E_{\lambda,n}^{\xi})\right]\label{eq:U}.
\eea
On the other hand, according to the basic thermodynamic relation $T dS = dU - \mu dN$ between entropy $S$, internal energy $U$ and particle number $N$, the net heat current transmitted across the junction arising from the node $\mathbf{K}_{\xi}$ (for $\xi=\pm$) is
\begin{equation}
\dot{Q}_{\xi} = \dot{U}_{\xi} -\left( \mu_L \dot{N}_{L}^{\xi}- \mu_R \dot{N}_{R}^{\xi}\right) .
\label{eq:heatcurrent}
\end{equation}
The thermal conductance is defined, as usual, under the condition that the net electric current vanishes ($I=0$)
\begin{eqnarray}
\kappa(T,V) = - \left.\frac{\partial \dot{Q}}{\partial \Delta T}\right|_{I = 0} = - \left.\frac{\partial \dot{U}}{\partial \Delta T}\right|_{I = 0},\label{eq:cond_therm}
\end{eqnarray}
where $\Delta T = T_R - T_L$ is the temperature difference between the contacts and  the total heat flux is given by the superposition from both Weyl nodes $\dot{Q} = \dot{Q}_{+} + \dot{Q}_{-}$, and similar relations hold for the total energy flux $\dot{U}$ and the total electric current $I$. The condition of a vanishing electric current defines an implicit relation between the voltage difference and the thermal gradient across the junction, by $I(\Delta T, V,T) = 0$. Therefore, we obtain the Seebeck coefficient by applying the implicit function theorem\cite{Munoz2019}
\begin{eqnarray}
S(T,V) = -\left.\frac{\partial V}{\partial \Delta T}\right|_{I=0,T}  =\frac{ \left.\displaystyle\frac{\partial I}{\partial \Delta T}\right|_{T,V}}{\left.\displaystyle\frac{\partial I}{\partial V}\right|_{T,\Delta T}},\label{eq:Seebeck}
\end{eqnarray}
where the temperature difference accross the junction $\Delta T(V,T)$ is obtained as the solution of the equation $I(T,V,\Delta T) = 0$. Following the argument above, the thermal conductance defined in Eq.\eqref{eq:cond_therm} is calculated by means of  the chain rule and in terms of the Seebeck coefficient\cite{Munoz2019}
\begin{equation}
\kappa(T,V) = -\left.\frac{\partial \dot{ U}}{\partial\Delta T}\right|_{T,V} + S(T,V)\left.\frac{\partial \dot{ U}}{\partial V}\right|_{ T,\Delta T}.\label{eq:thermal_conductance}
\end{equation}
From the general relations discussed above among the thermoelectric transport coefficients, we obtain the explicit formulae (see Sec. 4 of the Supplementary Material for mathematical details) for the thermal conductance
\bea
\kappa(T,V) &=&  \frac{v_F }{k_B (T + \Delta T)^2}\sum_{\xi,\lambda,n}\mathcal{T}(E_{\lambda,n}^{\xi})  E_{\lambda,n}^{\xi}\left[ E_{\lambda,n}^{\xi} - \mu  \right] f_R(E_{\lambda,n}^{\xi})\left[
1 - f_R(E_{\lambda,n}^{\xi})\right] \nonumber\\
& &\quad+\,\, S(T,V)\frac{e v_F}{k_B T} \sum_{\lambda,n,\xi}\mathcal{T}(E_{\lambda,n}^{\xi})E_{\lambda,n}^{\xi}\,f_L(E_{\lambda,n}^{\xi}) \left[ 1-f_L(E_{\lambda,n}^{\xi})  \right], \label{eq:kappa_final_mag_torsion_RDSP}
\eea
and for the Seebeck coefficient
\bea
S(T,V) =-\frac{T\sum_{\lambda,n,\xi}\mathcal{T}(E_{\lambda,n}^{\xi})\left(E_{\lambda,n}^{\xi} - \mu\right) f_R(E_{\lambda,n}^{\xi})\left[1- f_R(E_{\lambda,n}^{\xi}) \right]}
{e (T + \Delta T)^2\sum_{\lambda,n,\xi}\mathcal{T}(E_{\lambda,n}^{\xi})f_L(E_{\lambda,n}^{\xi})\left[1- f_L(E_{\lambda,n}^{\xi}) \right]}.
\label{eq:Seebeck_mag_torsion_RDSP}
\eea

%%%%%%%%%%%%%%%%%%%%%%%%%%%%%%%%%%%%%%%%%%
\section{Results}

\label{sec:Results}

In this section we will apply the analytical results derived in Section Theory to study the response of the transport coefficients to the relevant physical parameters of the model, such as the external magnetic field $B_0$, the torsion angle $\theta$, the temperature $T$, and the applied bias voltage $V$ \cite{Soto_Garrido_2018,Munoz2019}. In particular, we will analyze the effect of the RDSP, as a model for the lattice mismatch, by varying the $V_0$ parameter that characterizes the strength of the repulsive barrier, expressed in terms of the ``spinor rotation'' angle $\alpha = V_0/\hbar v_F$. \\

By considering first the case where only the lattice mismatch effect is present (RDSP) (see {\color{black}Eq.~(29) in the Sec. 2 of the Supplementary Material}), we notice that the phase shifts depend on the parameter $V_0$  through $\tan \alpha$. Therefore, the results depend on $\alpha$ periodically, with period $\pi$, as seen in Fig.~\ref{fig:tan_phase_shift_delta}. It is also clear (from {\color{black}Eq.~(29) in the Sec. 2 of the Supplementary Material}) that if the only scattering mechanism is the RDSP, the transmission is maximum for $\alpha=n\pi$, with $n$ an integer. At these particular ``magic'' values, despite the presence of the lattice mismatch, the corresponding interfacial energy barrier becomes transparent to the Weyl fermions of both chiralities $\xi=\pm$.  

In order to study the additional effect of torsion and magnetic field for TaAs, we estimate \cite{Pikulin2016} $B_S\approx1.8\times 10^{-3}T$ per angular degree of torsion. Furthermore, we have that the modified flux quantum in this material is approximately $\displaystyle\tilde{\phi}_{0}\equiv \frac{\hbar v_F}{e}=\frac{1}{2\pi}\frac{v_F}{c}\frac{hc}{e}=\frac{1}{2\pi}\frac{1.5}{300}\cdot4.14\times10^5$ T$\mathring{\text{A}}^2\approx330$ T$\mathring{\text{A}}^2$. Using these values, we obtain the simple relation between the torsional angle $\theta$ (in degrees) and the pseudo-magnetic field $B_S$ representing strain
\begin{equation}
B_S a^2=1.36\,\theta\, \tilde{\phi}_{0}. \label{eq:torsion_angle}
\end{equation}

In this case, the analytical expression for the scattering phase shift is given by {\color{black}Eq.~(32) in the Sec. 2 of the Supplementary Material}. We notice that the effect of the barrier is again given by $\tan\alpha$, and hence it becomes minimal at ``magic'' values of $\alpha = n\pi$, i.e. integer multiples of $\pi$. However, in this second case the scattering phase-shift does not vanish, due to the residual combined effect of torsion and magnetic field. This can be seen in Fig.~\ref{fig:tan_phase_shift_delta_mag}, where for $\alpha=0,\,\pi,\,2\pi$, $\tan \delta_m\neq0$, in contrast to Fig.~\ref{fig:tan_phase_shift_delta}. Actually, the value of $\tan \delta_m$ for $\alpha=n\pi$ ($n$ integer) and the consequences of the scattering by the combined magnetic field and torsion, but in the absence of the lattice mismatch barrier, was extensively discussed in our previous works\cite{Munoz_2017,Soto_Garrido_2018,Munoz2019}. 

Another important aspect to notice is that, when we only consider the lattice mismatch effect, the scattering phase shift is symmetric for both chiral nodes $\xi = \pm 1$, as seen in Subfigs. (a) and (b) in Fig.~\ref{fig:tan_phase_shift_delta}. In contrast, when the magnetic field and torsion are present, this symmetry is broken, as displayed in Subfigs. (a) and (b) in Fig.~\ref{fig:tan_phase_shift_delta_mag}. As we explained in Ref.~\cite{Soto_Garrido_2018}, this occurs because the magnitude of the pseudo-field that combines torsion and magnetic field $B_{\xi} = B_0 + \xi B_S$ depends on the sign of the node chirality, a manifestation of the chiral anomaly which can be also observed in the electric current (see Fig.~\ref{fig:node_conductance_a}). \\

\begin{figure}
\centering
\subfigure[$\xi=+1$]{\includegraphics[width=0.48\textwidth]{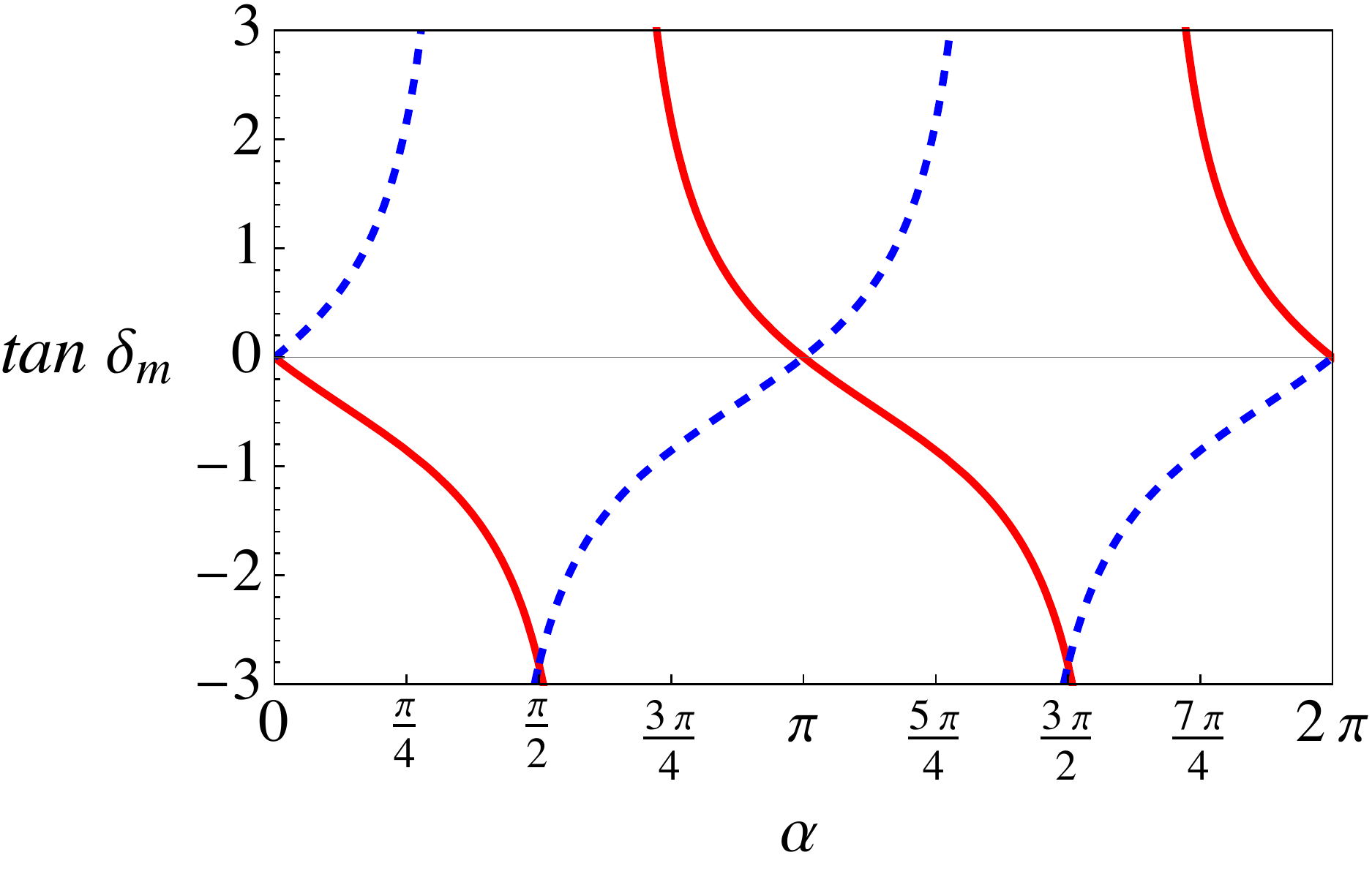}\label{fig:tan_phase_shift_delta_a} }
\hskip .1cm 
\subfigure[$\xi=-1$]{\includegraphics[width=0.48\textwidth]{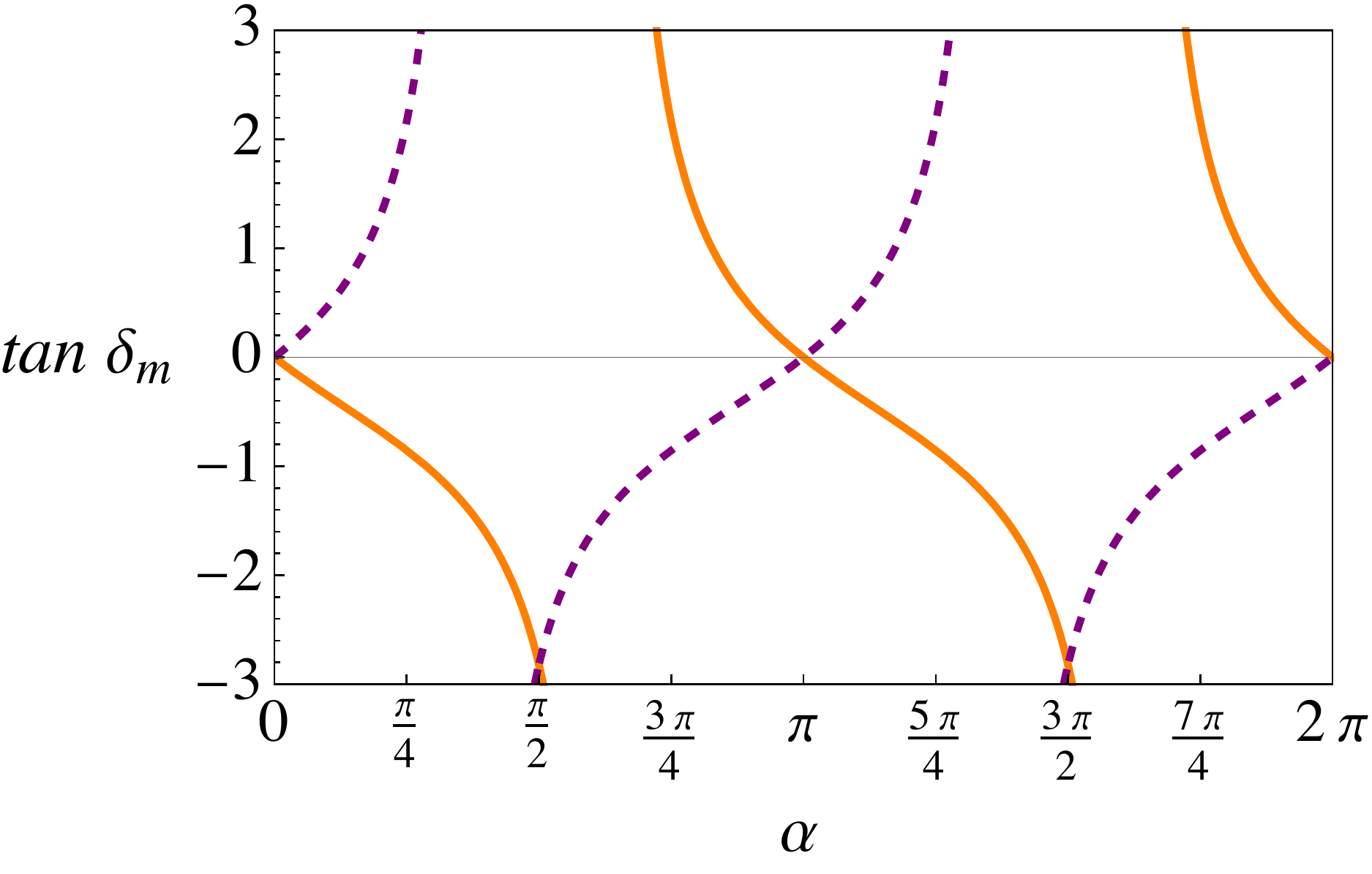}\label{fig:tan_phase_shift_delta_b}}
\caption{(Color online) Analytical expression for $\tan \delta_m$ (in Eq.(29) in Supporting Information) plotted as a function of $\alpha$. The plots are computed for a wave vector $k_{\perp}\sim1/a$ and an orbital angular momentum $m=1$. The subfig. (a) is for a node index $\xi=1$; the red (solid) line corresponds to a band index $\lambda=1$ and the blue (dashed) line is for $\lambda=-1$. The subfig. (b) is for a node index $\xi=-1$; the orange (solid) line corresponds to a band index $\lambda=1$ and the purple (dashed) line is for $\lambda=-1$.}
\label{fig:tan_phase_shift_delta}
\end{figure}

\begin{figure}
\centering
\subfigure[$\xi=-1$]{\includegraphics[width=0.48\textwidth]{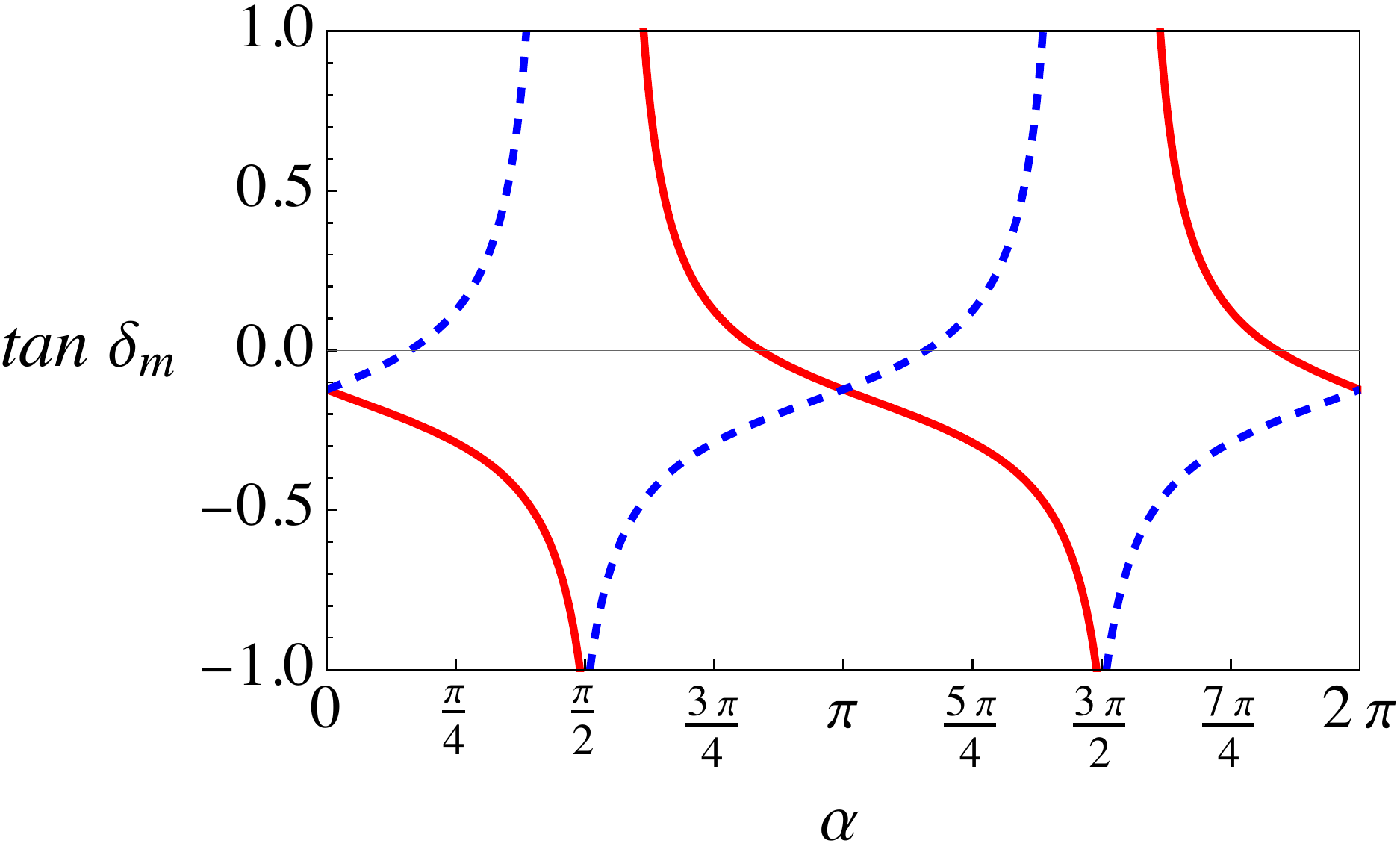}\label{fig:tan_phase_shift_delta_mag_a} }
\hskip .1cm 
\subfigure[$\xi=+1$]{\includegraphics[width=0.48\textwidth]{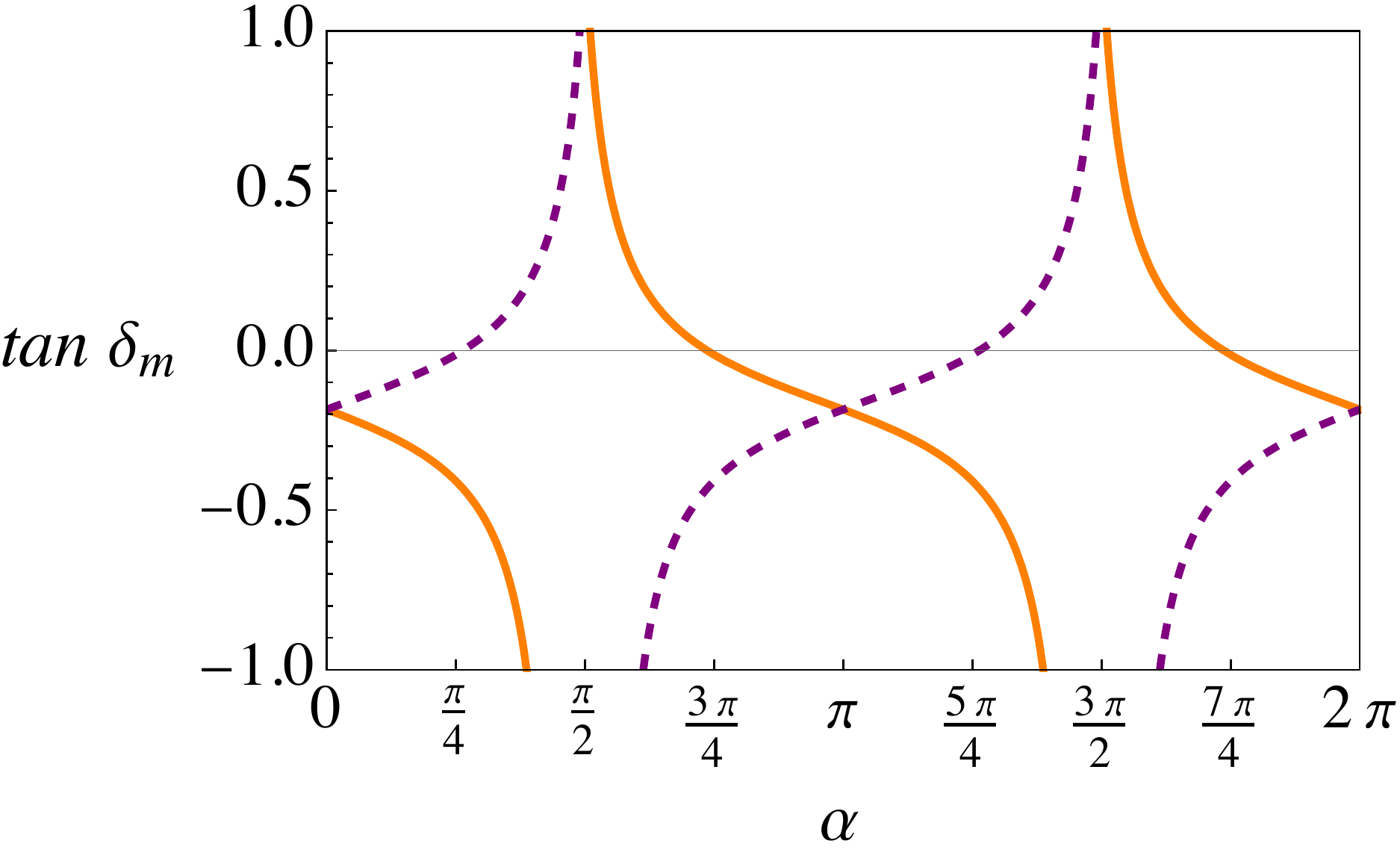}\label{fig:tan_phase_shift_delta_mag_b}}
\caption{(Color online) Analytical expression for $\tan \delta_m$ (in Eq.(32) of Supporting Information) plotted as a function of $\alpha$. The plots are computed for a quantum number $n=1$, orbital angular momentum $m=1$, an external magnetic field $B_0 a^2=25\tilde{\phi}_0$ and a torsion angle $\theta=10^{\circ}$. The Subfig. (a) is for a node index $\xi=1$; the red (solid) line corresponds to a band index $\lambda=1$ and the blue (dashed) line is for $\lambda=-1$. The Subfig. (b) is for a node index $\xi=-1$; the orange (solid) line corresponds to a band index $\lambda=1$ and the purple (dashed) line is for $\lambda=-1$.   }
\label{fig:tan_phase_shift_delta_mag}
\end{figure}

\subsection{Electronic transport}
\label{subsec:electronic_transport}

\begin{figure}
	\centering
\subfigure[]{\includegraphics[width=0.48\textwidth]{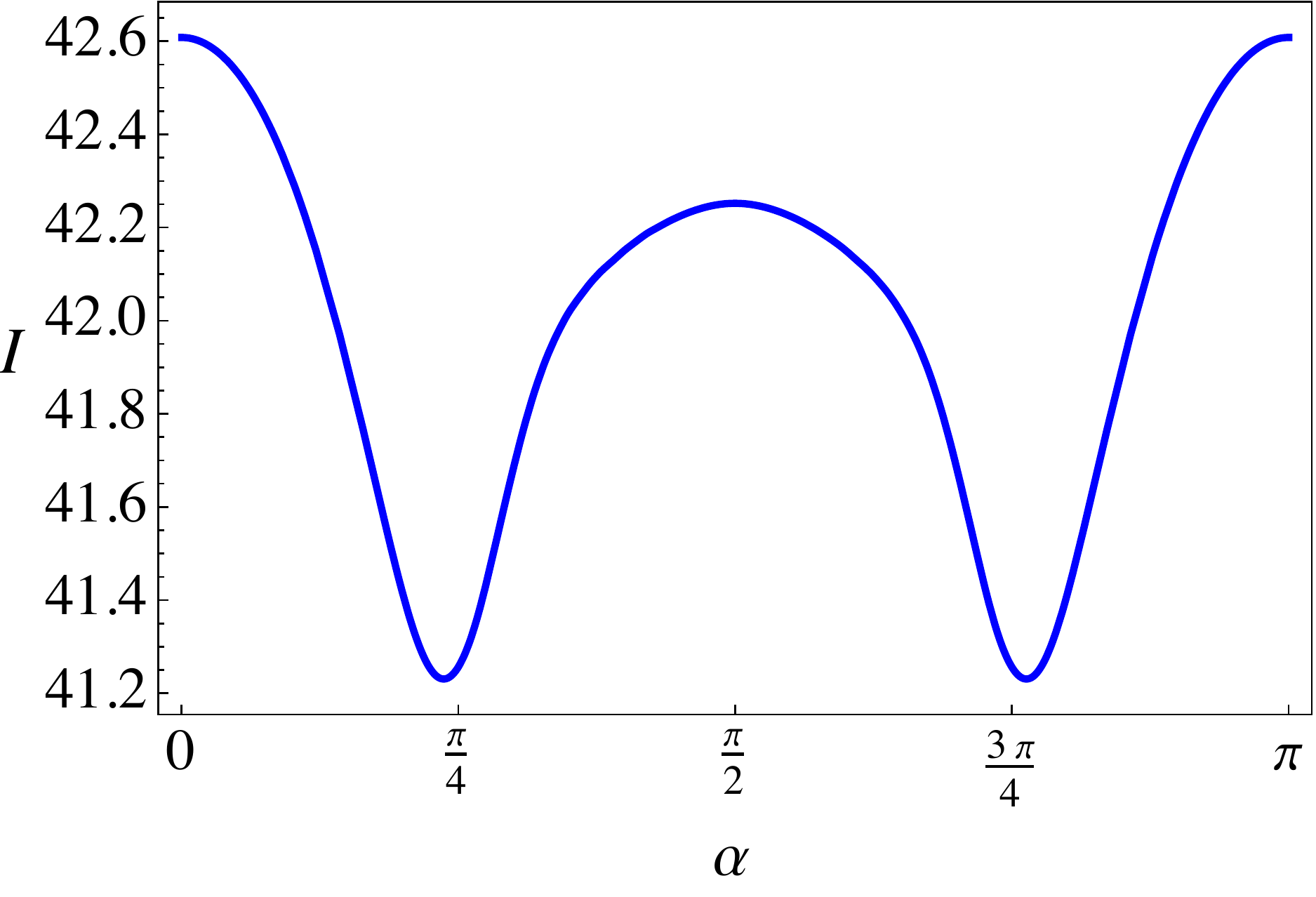}\label{fig:current_delta_a} }
\subfigure[]{\includegraphics[width=0.48\textwidth]{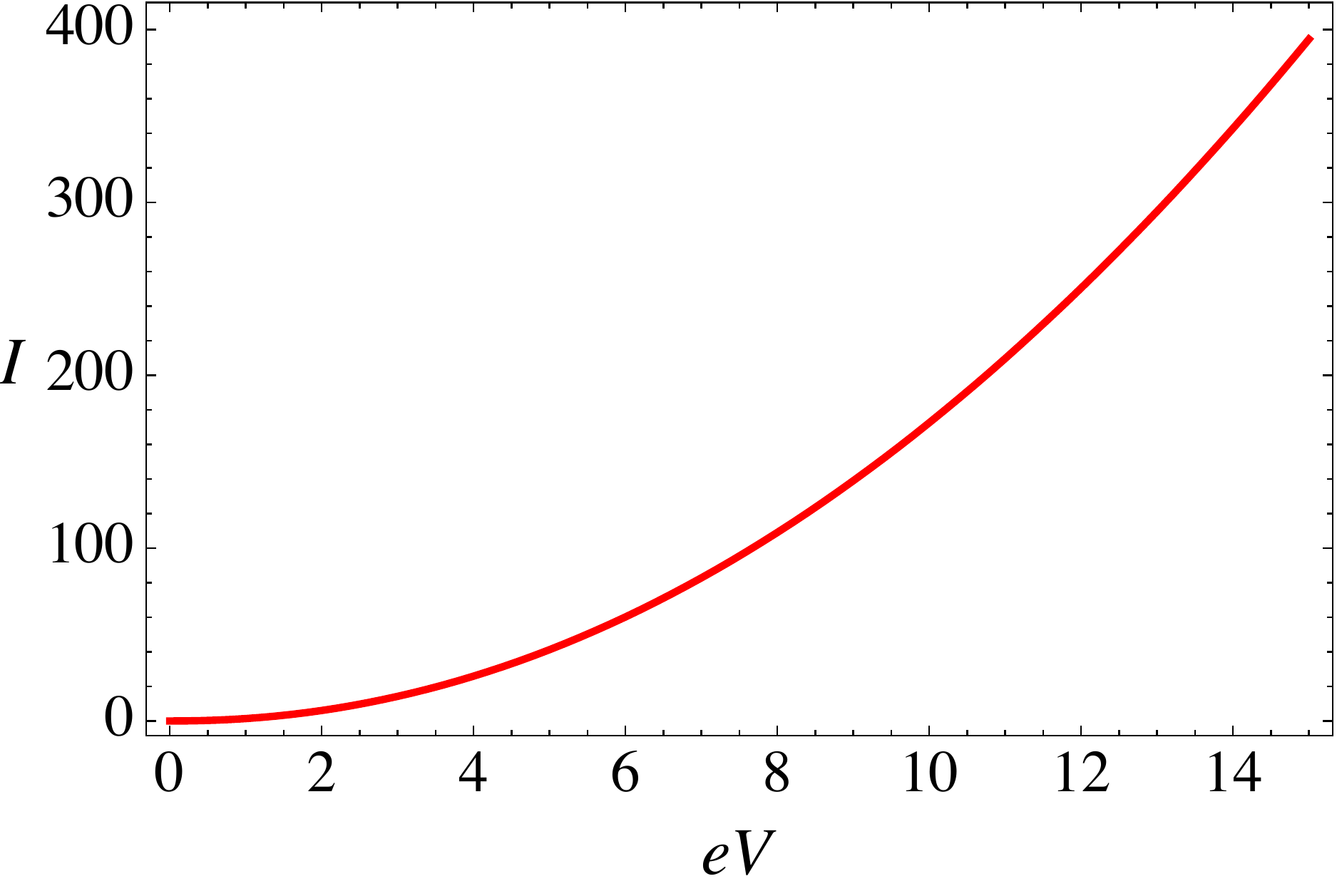}\label{fig:current_delta_b} }
	\hskip .1cm 
	\caption{(Color online)  Electric current (in units of $ev_F/a$) computed from the analytical expression in Eq.~\eqref{eq:node_elec_current_transm_delta} for the case of the RDSP barrier alone and $T=0.2\,\hbar v_F/k_B a$: (a) Plotted as a function of the applied bias $eV$ (in units of $\hbar  v_F/a$)  and (b)  plotted as a function of $\alpha$ (dimensionless). }
	\label{fig:current_delta}
\end{figure}

\begin{figure}
	\centering
\includegraphics[width=0.48\textwidth]{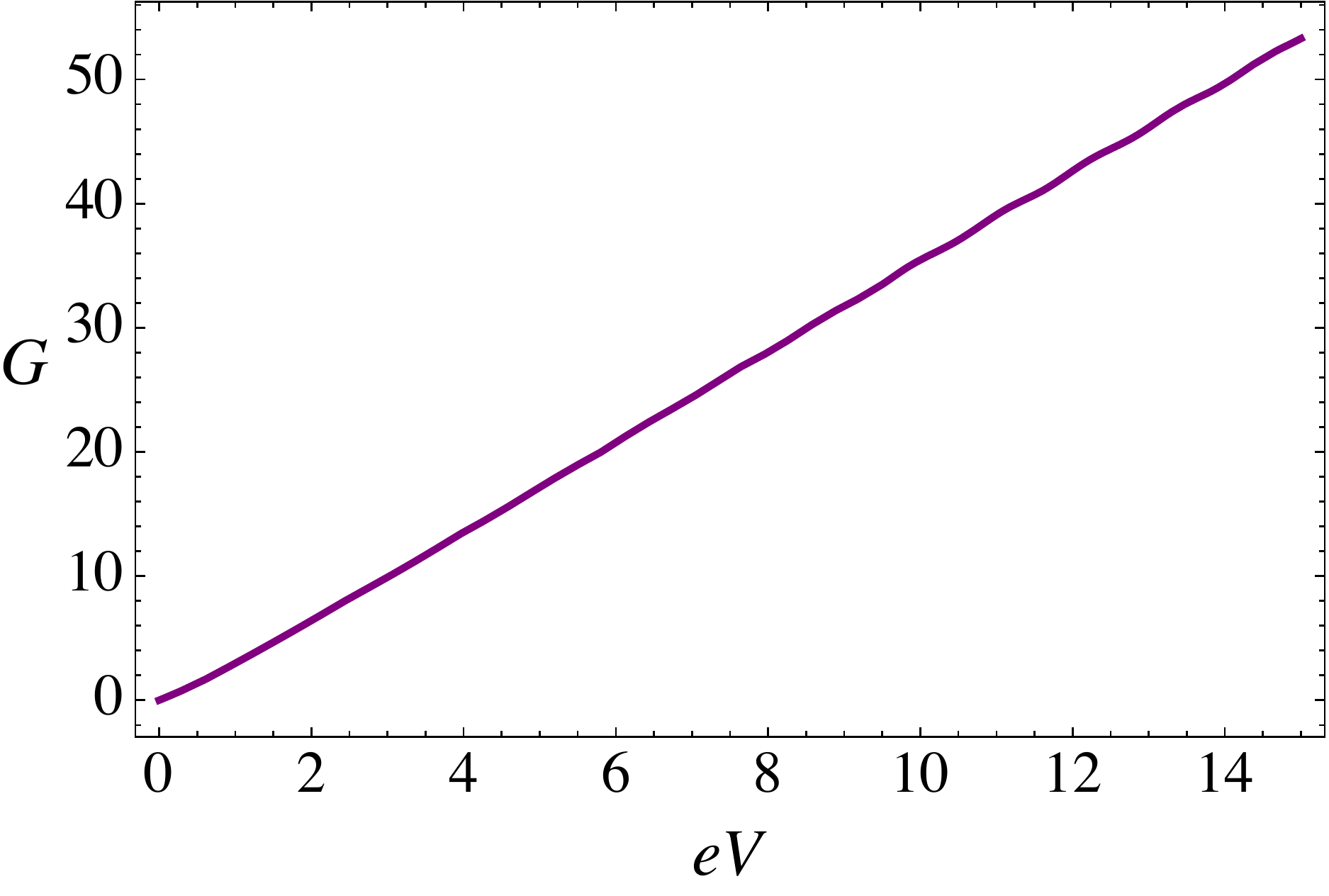}
	\hskip .1cm 
	\caption{(Color online) Differential conductance (in units of $e^2/\hbar$), for the RDSP barrier alone, plotted as function of applied bias $eV$ (in units of $\hbar  v_F/a$) for $\alpha=3\pi/4$ and $T=0.2\,\hbar v_F/k_B a$, computed from Eq.~\eqref{eq:conductance_delta}.}
	\label{fig:conductance_delta}
\end{figure}

The electric current (in units of $ev_F/a$) is computed from Eq.~\eqref{eq:node_elec_current_transm_delta} for the case of the RSDP only, in the absence of torsion and magnetic field. Fig.~\ref{fig:current_delta_a} shows the periodic dependence of the total current as a function of the dimensionless parameter $\alpha = V_0/(\hbar v_F)$ that characterizes the magnitude of the lattice mismatch barrier, and for a temperature $T=0.2\,\hbar v_F/k_B a$. As expected, the maxima of transmission occur for the ``magic angles'' $\alpha = n \pi$ ($n$ integer), and the overall effect of the barrier is to slightly reduce the current, reaching minimal values near $\alpha = \pi/4$ and $\alpha = 3\pi/4$, respectively. The behavior of the current, for the same temperature, as a function of the bias voltage is presented in Fig.~\ref{fig:current_delta_b}. As can be seen, for low temperatures the current across the junction displays an approximately quadratic dependence on the applied bias voltage $eV$ (in units of $\hbar  v_F/a$), that leads to an approximately linear dependence of the differential conductance (in units of $e^2/\hbar$) on the bias voltage in Fig.~\ref{fig:conductance_delta}. 

\begin{figure}
	\centering
	\subfigure[]{\includegraphics[width=0.48\textwidth]{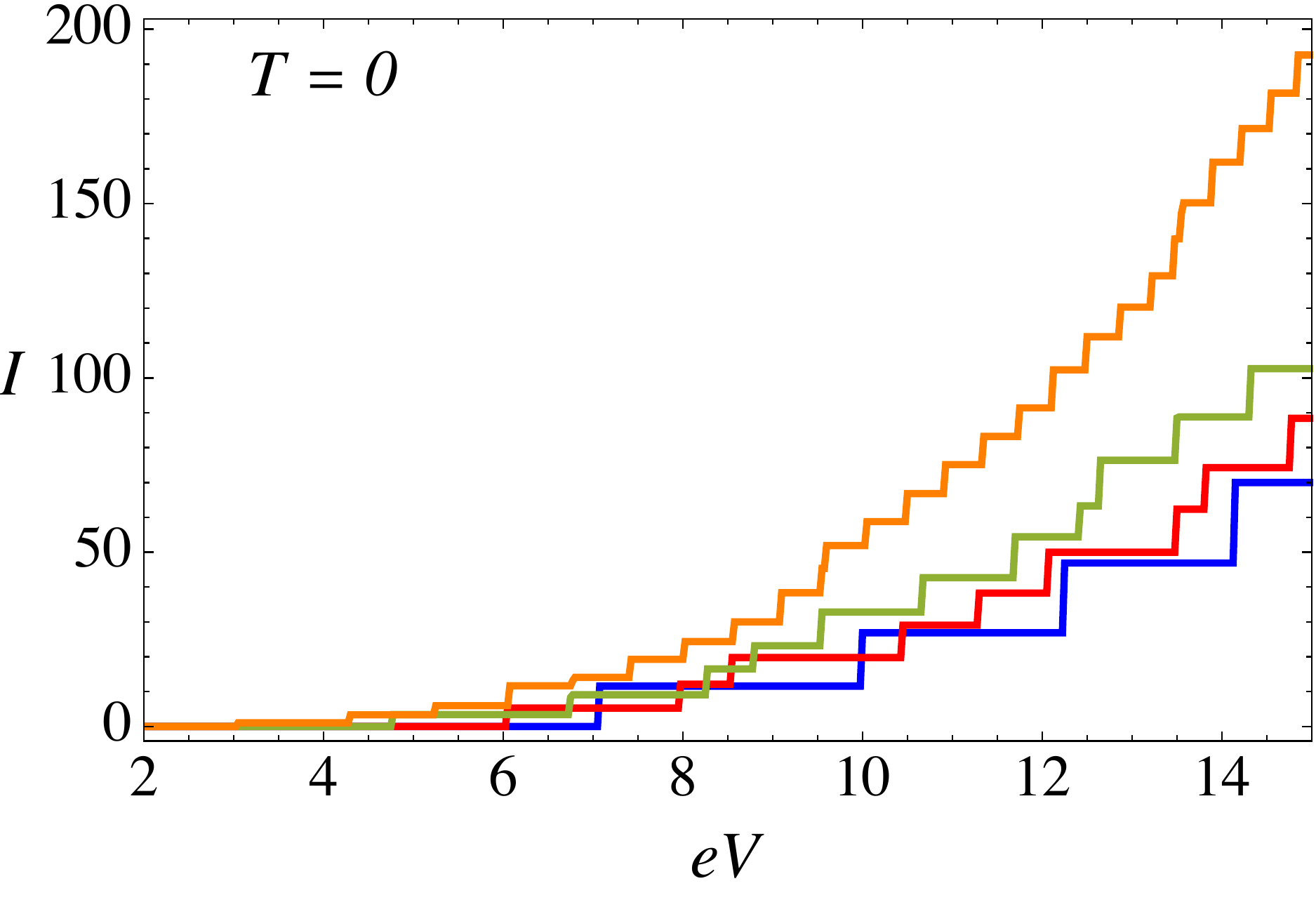}\label{fig:currentsT0_a} }
	\hskip .1cm 
	\subfigure[]{\includegraphics[width=0.48\textwidth]{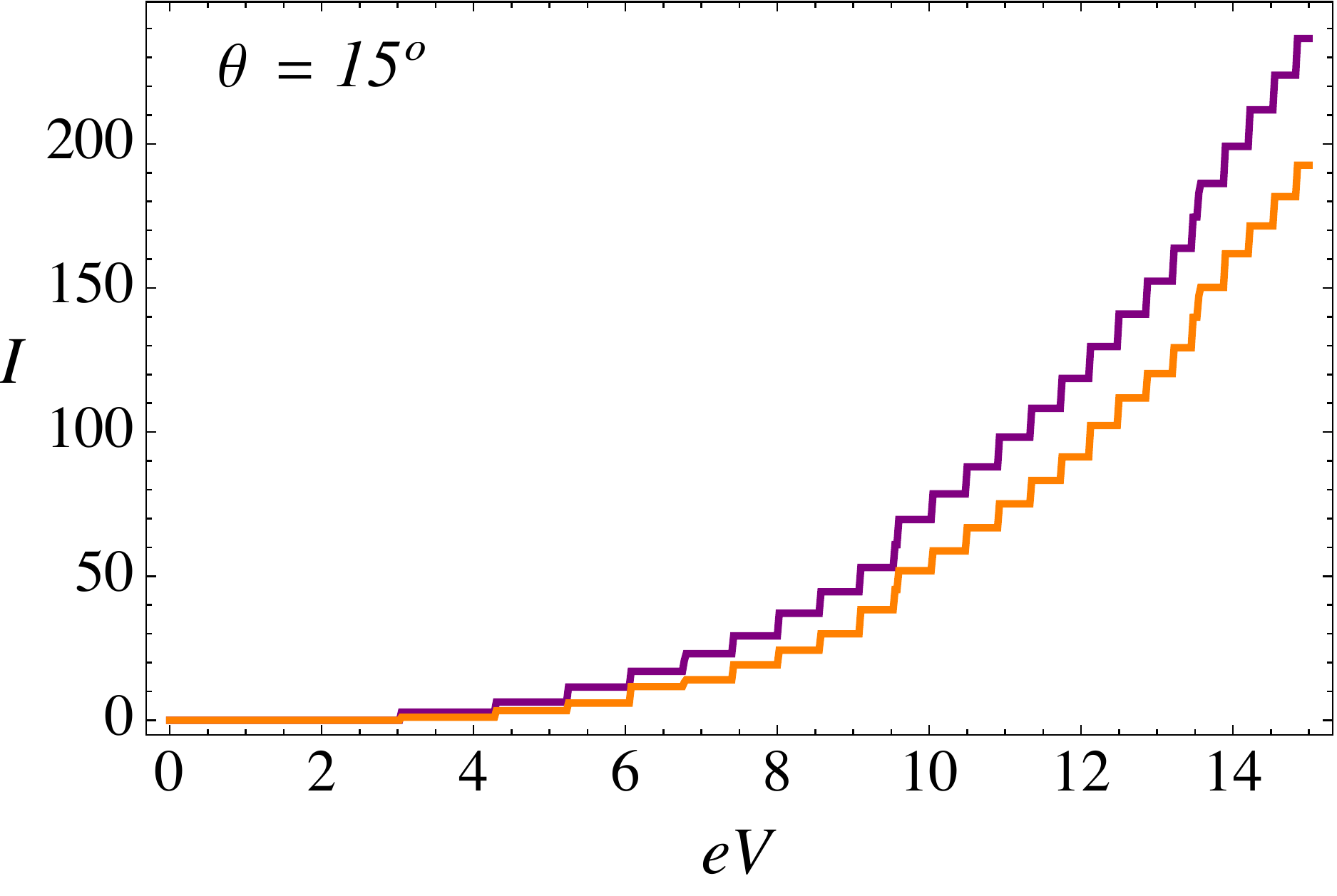}\label{fig:currentsT0_b}}
	\caption{(Color online) (a) Electric current (in units of $ev_F/a$) as function of applied bias $eV$ (in units of $\hbar  v_F/a$), computed from the analytical expression in Eq.~\eqref{eq:node_elec_current_transm_mag_torsion_RDSP} at zero temperature, for an external magnetic field $B_0 a^2=25 \tilde{\phi}_0$ and $\alpha=3\pi/4$. The blue line corresponds to a twist angle $\theta=0^{\circ}$, red is for $\theta=5^{\circ}$, green is for $\theta=10^{\circ}$ and the orange line corresponds to $\theta=15^{\circ}$. (b) Comparison of electric currents at zero temperature, for $B_0 a^2=25 \tilde{\phi}_0$ and $\theta=15^{\circ}$: the purple line is for $\alpha=0$, and the orange line corresponds to $\alpha=3\pi/4$.}
	\label{fig:currentsT0}
\end{figure}

\begin{figure}
	\centering
	\subfigure[]{\includegraphics[width=0.48\textwidth]{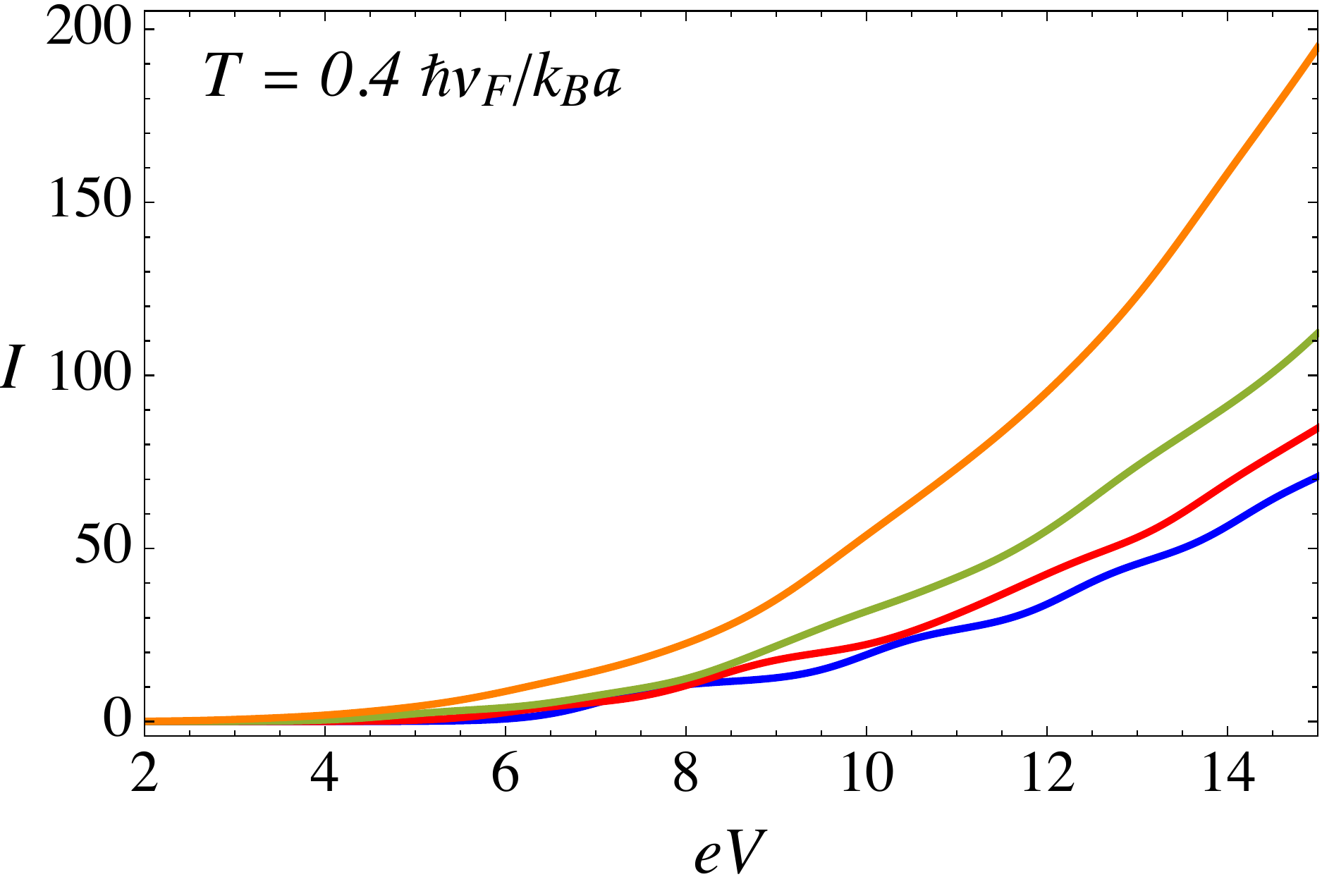}\label{fig:currentsTfinite_a} }
	\hskip .1cm 
	\subfigure[]{\includegraphics[width=0.48\textwidth]{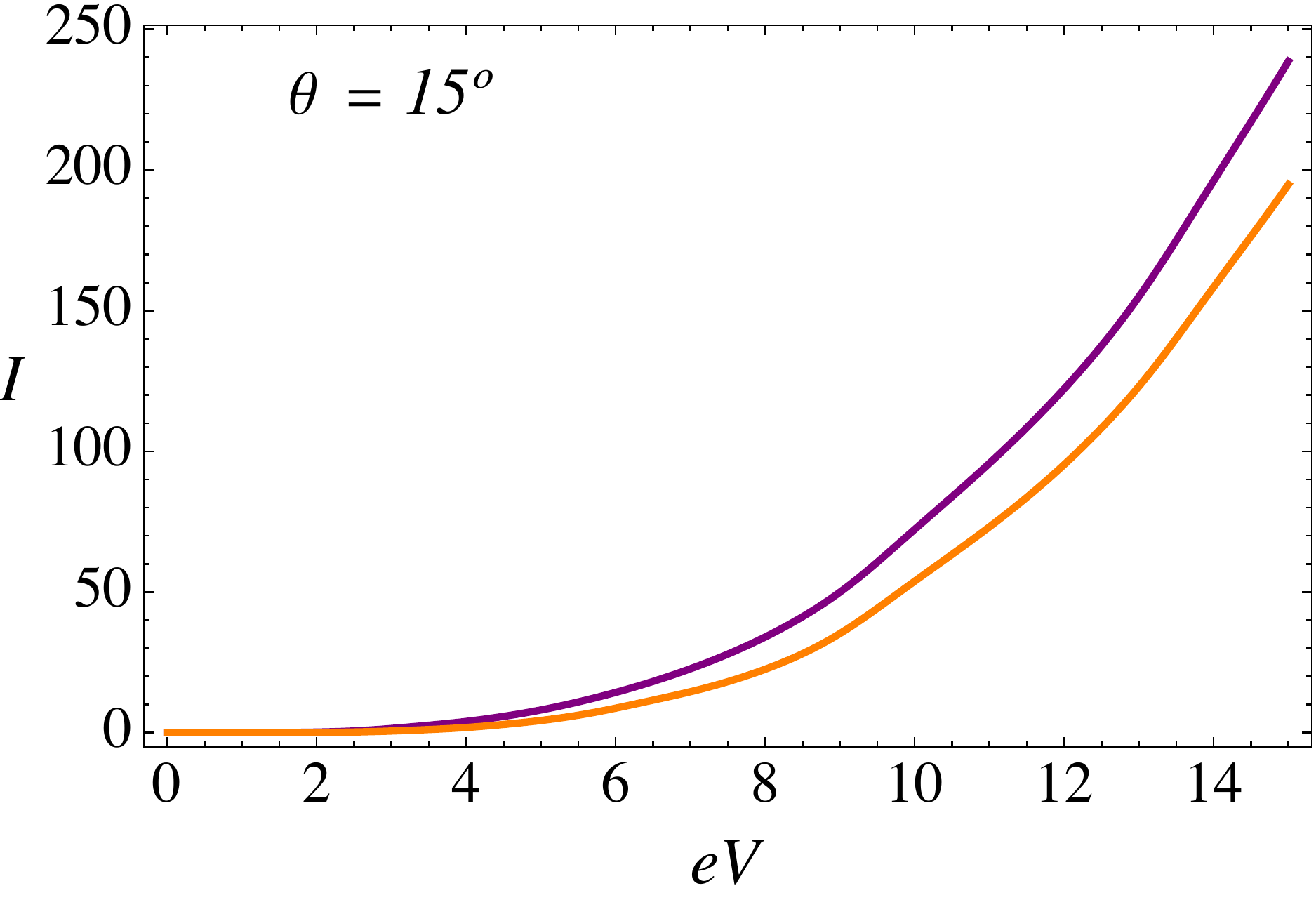}\label{fig:currentsTfinite_b}}
	\caption{(Color online) (a) Electric current (in units of $ev_F/a$) plotted as function of applied bias $eV$ (in units of $\hbar  v_F/a$), computed from the analytical expression in Eq.~\eqref{eq:node_elec_current_transm_mag_torsion_RDSP} at  $T=0.4\,\hbar v_F/k_B a$, for an external magnetic field $B_0 a^2=25 \tilde{\phi}_0$ and $\alpha=3\pi/4$. The blue line corresponds to a twist angle $\theta=0^{\circ}$, red is for $\theta=5^{\circ}$, green is for $\theta=10^{\circ}$ and the orange line corresponds to $\theta=15^{\circ}$. (b) Comparison of electric currents at $T=0.4\,\hbar v_F/k_B a$, for $B_0 a^2=25 \tilde{\phi}_0$ and $\theta=15^{\circ}$: the purple line is for $\alpha=0$ and orange corresponds to $\alpha=3\pi/4$.}
	\label{fig:currentsTfinite}
\end{figure}

\begin{figure}
	\centering
	\subfigure[]{\includegraphics[width=0.48\textwidth]{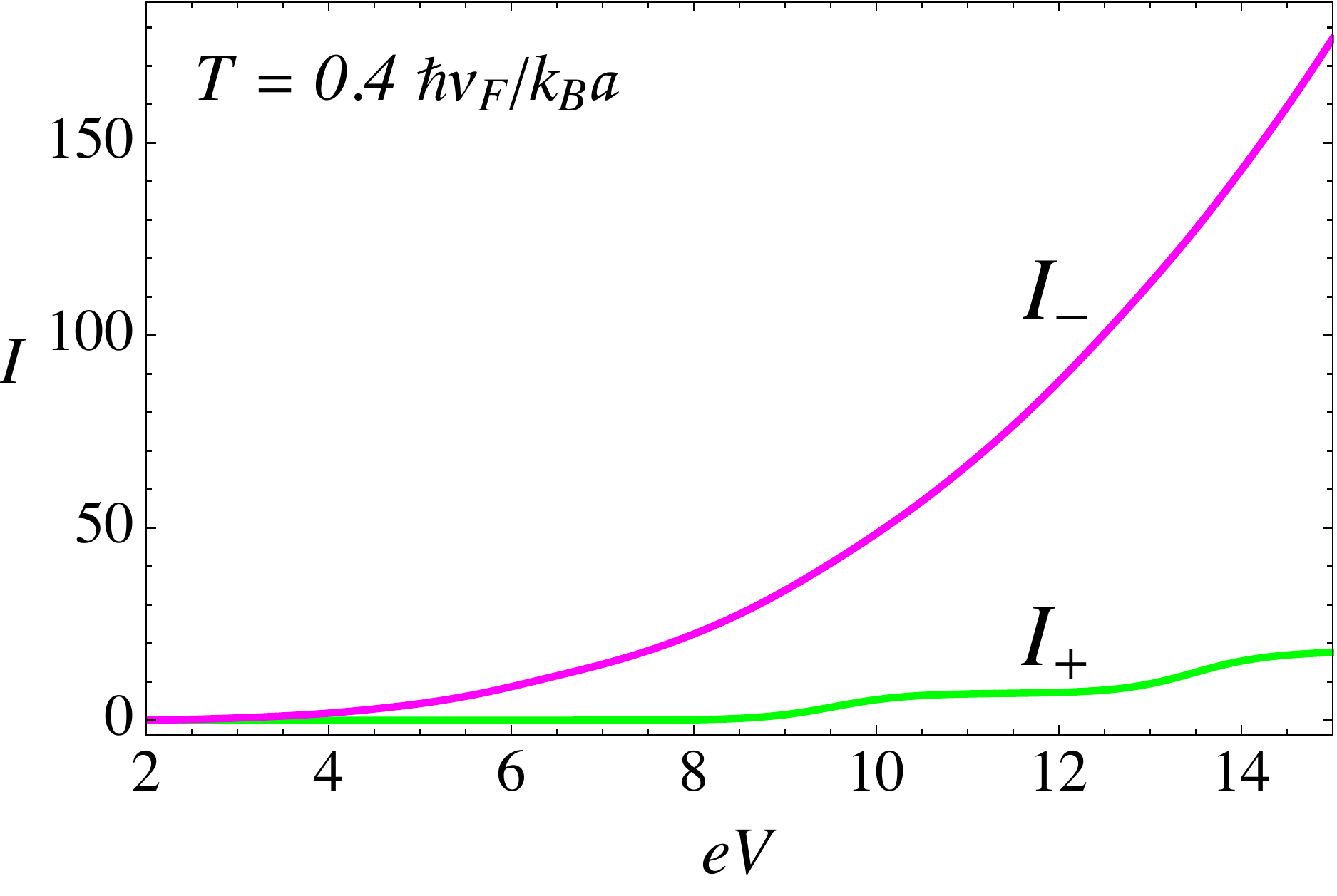}\label{fig:node_conductance_a} }
	\hskip .1cm 
	\subfigure[]{\includegraphics[width=0.48\textwidth]{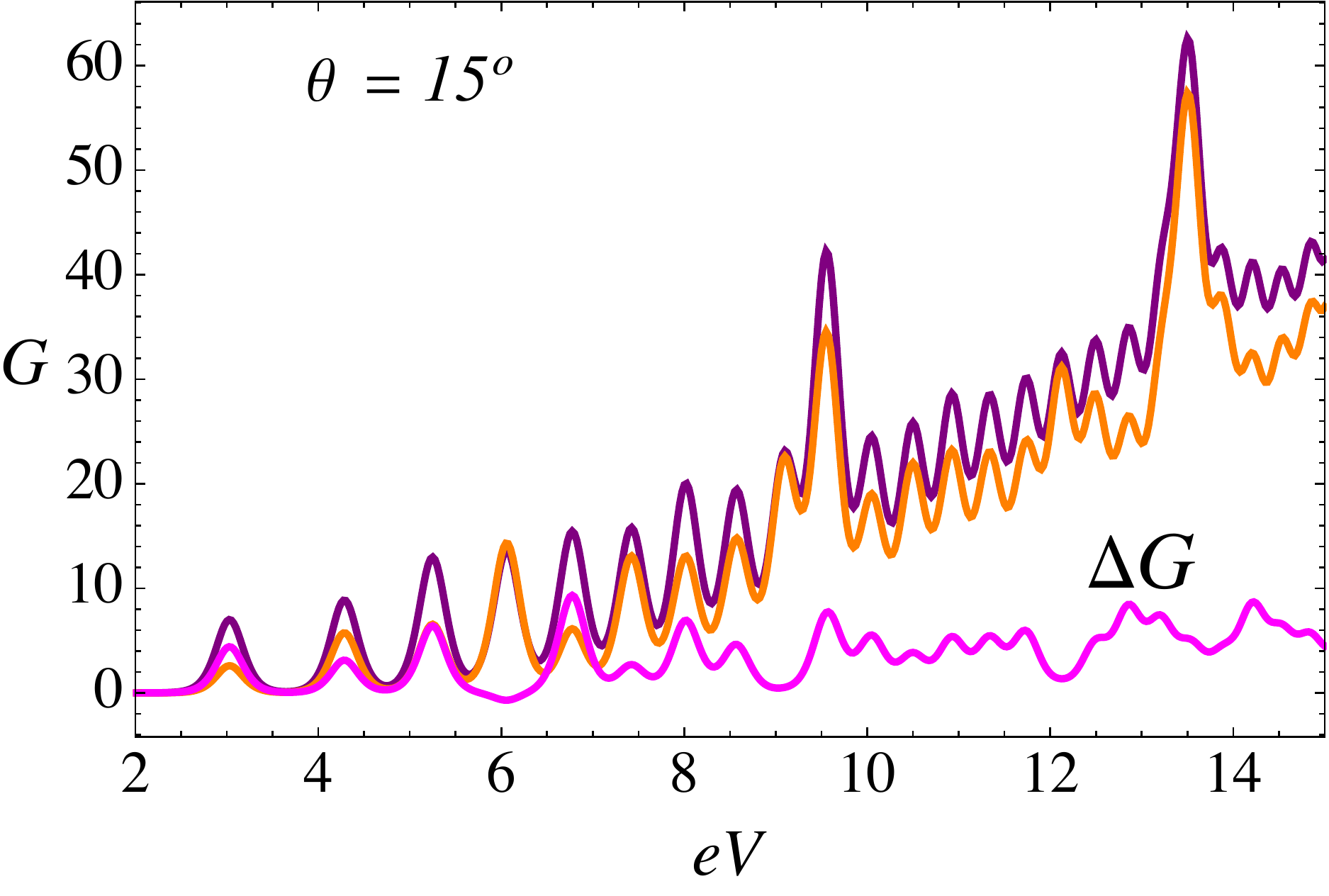}\label{fig:node_conductance_b}}
	\caption{(Color online) (a) Node-polarized components of the currents computed for an external magnetic field $B_0 a^2=25\tilde{\phi}_0$, a torsion angle $\theta=15^{\circ}$ and $\alpha=3\pi/4$: the magenta line corresponds to the contribution of $I_{-}$ arising from the $\mathbf{K}_{-}$ node and the green line corresponds to the contribution of $I_{+}$ arising from the $\mathbf{K}_{+}$ node. (b) Comparison of conductance (in units of $e^2/\hbar$) as a function of the bias voltage $eV$ (in units of $\hbar  v_F/a$) for the case of an external magnetic field $B_0 a^2=25 \tilde{\phi}_0$, $T=0.1\hbar v_F/k_B a$, and a torsion angle $\theta=15^{\circ}$.  The purple line is for $\alpha=0$, orange corresponds to $\alpha=3\pi/4$ and the magenta line corresponds to the difference between both $\Delta G=G(\alpha=0)-G(\alpha=3\pi/4)$.      }
	\label{fig:node_conductance}
\end{figure}
Now, when we include the combined effect of the delta barrier, the external magnetic field, and the torsion strain, the current is calculated from the analytical expression in Eq.~\eqref{eq:node_elec_current_transm_mag_torsion_RDSP}. Fig.~\ref{fig:currentsT0_a} presents the total current as a function of voltage at zero temperature, an external field $B_0 a^2=25 \tilde{\phi}_0$,
a value of $\alpha=3\pi/4$, and different values of the torsion angle $\theta$. A remarkable feature at zero temperature is the appearance of plateaus in the current; the elastic scattering condition explains it because the incident particle energy must be resonant to one of the pseudo-Landau levels inside the cylinder, and hence each subsequent plateau corresponds to the transmission of an additional Landau level. Such plateaus tend to be smoothed with increasing temperature, as can be seen in Fig.~\ref{fig:currentsTfinite_a}. As we discussed in our previous work in the absence of the RDSP contribution\cite{Soto_Garrido_2018,Munoz2019}, for a fixed external magnetic field the electric current increases with the torsion angle $\theta$. This effect is due to an enhanced transmission of the Weyl fermions arising from the $\mathbf{K}_{-}$ node, since for this particular chirality $\xi=-1$ the magnitude of the effective pseudo-magnetic field $|B^{-}|=|B_0-B_S|$ is smaller, thus increasing the spectral density of pseudo-Landau levels ($\sim\sqrt{|B^{\xi}|n}$) for chirality $\xi=-1$, and consequently an increase in the number of channels available for transmission.  Fig.~\ref{fig:node_conductance_a}  presents the difference between the currents originated at each node. Furthermore, for a fixed torsion angle, the transmitted current decreases as the external magnetic field increases\cite{Soto_Garrido_2018}. This effect occurs because, by increasing the external field $B_0$ (for a fixed torsion field $B_S$), the magnitude of the effective pseudo-magnetic field $|B^{\xi}|=|B_0 + \xi B_S|$ increases for both chiralities $\xi=\pm$, thus reducing the density of Landau levels available for transmission. Fig.~\ref{fig:currentsT0_b} and Fig.~\ref{fig:currentsTfinite_b} present a comparison of the current, for $\alpha=0$ and $\alpha=3\pi/4$, at $T=0$ and $T=0.4\,\hbar v_F/k_B a$, respectively. We see that the magnitude of the current is reduced while the position of the plateaus remains fixed. This effect is more significant at higher bias voltage, and is due to the repulsive effect of the RDSP barrier that reduces the transmission. Finally, Fig.~\ref{fig:node_conductance_b} compares the conductance (in units of $e^2/\hbar$) as a function of the bias voltage $eV$ (in units of $\hbar  v_F/a$) for the case of an external magnetic field $B_0 a^2=25 \tilde{\phi}_0$, a torsion angle $\theta=15^{\circ}$,  $T=0.1\hbar v_F/k_B a$, and two different values of the lattice mismatch RDSP barrier, $\alpha= 0$ and $\alpha = 3\pi/4$, respectively. As expected, the conductance shows peaks as a consequence of the plateaus observed in the current. The effect of the RDSP barrier is to reduce the conductance without affecting the position of the peaks.

\subsection{Thermal transport}
\label{subsec:thermal_transport}

\begin{figure}
	\centering
	\subfigure[]{\includegraphics[width=0.48\textwidth]{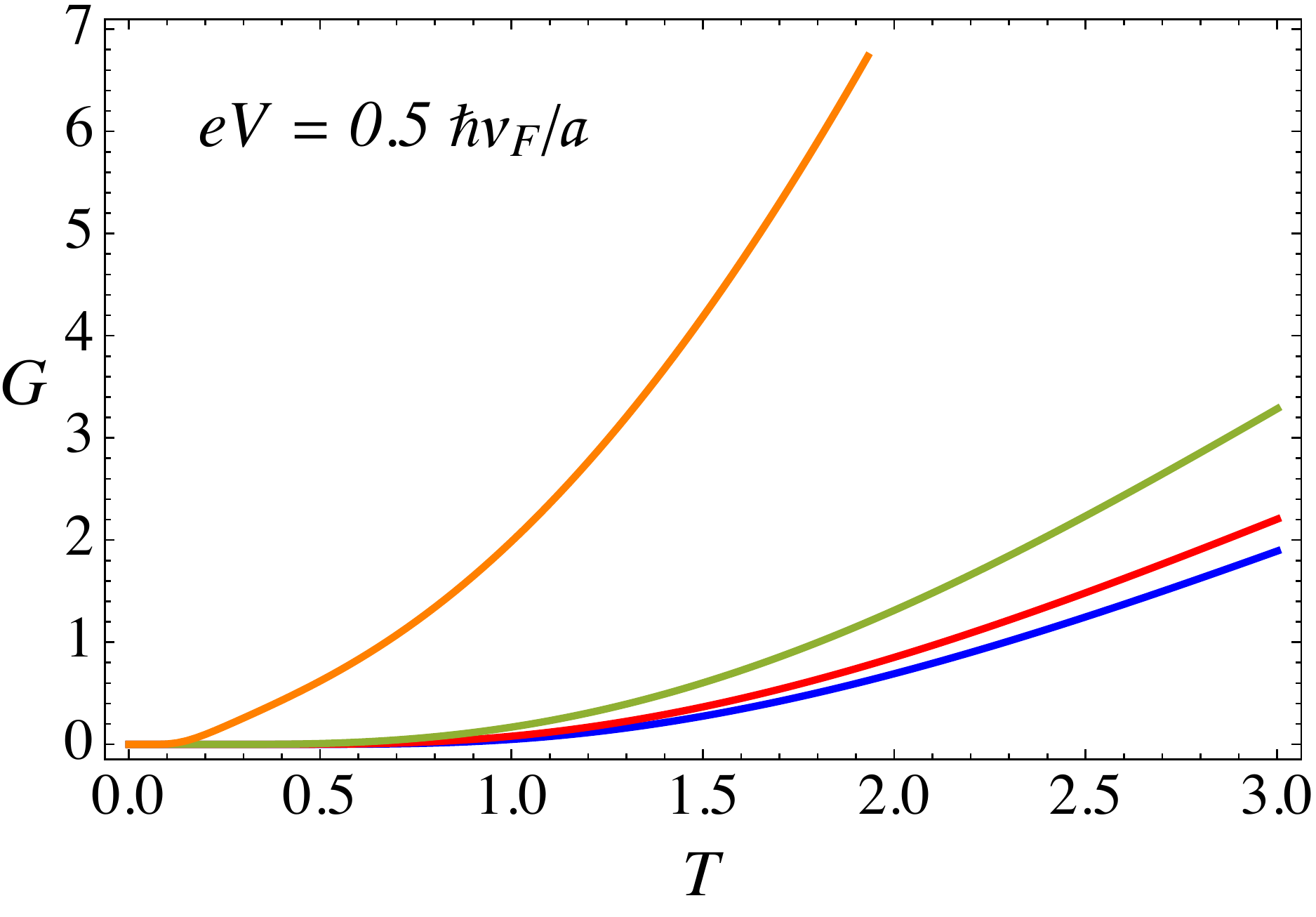}\label{fig:conductanceT_a} }
	\hskip .1cm 
	\subfigure[]{\includegraphics[width=0.48\textwidth]{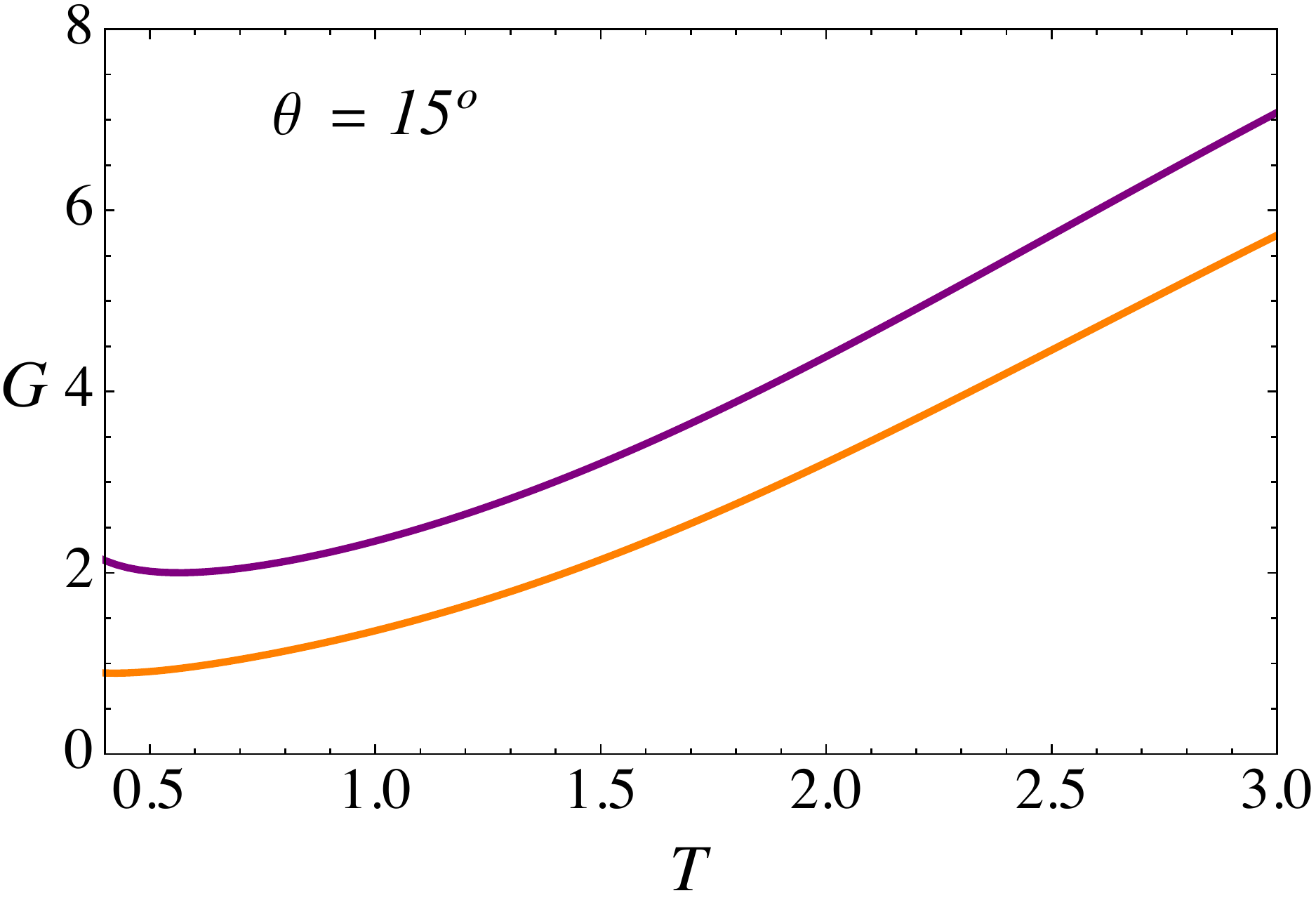}\label{fig:conductanceT_b}}
	\caption{(Color online) (a) Conductance (in units of $e^2/\hbar$) as a function of temperature (in units of $\hbar v_F/k_B a$) for external $B_0a^2=25\tilde{\phi}_0$, a bias $e V = 0.5\, \hbar v_F/a$ and $\alpha=3\pi/4$. The blue line corresponds to $\theta=0^{\circ}$, red is for $\theta=5^{\circ}$, green is for $\theta=10^{\circ}$ and the orange line corresponds to $\theta=15^{\circ}$. (b) Comparison of conductance for $\theta=15^{\circ}$: the purple line is for $\alpha=0$ whereas the orange line is for $\alpha=3\pi/4$.}
	\label{fig:conductanceT}
\end{figure}

\begin{figure}
	\centering
	\subfigure[]{\includegraphics[width=0.48\textwidth]{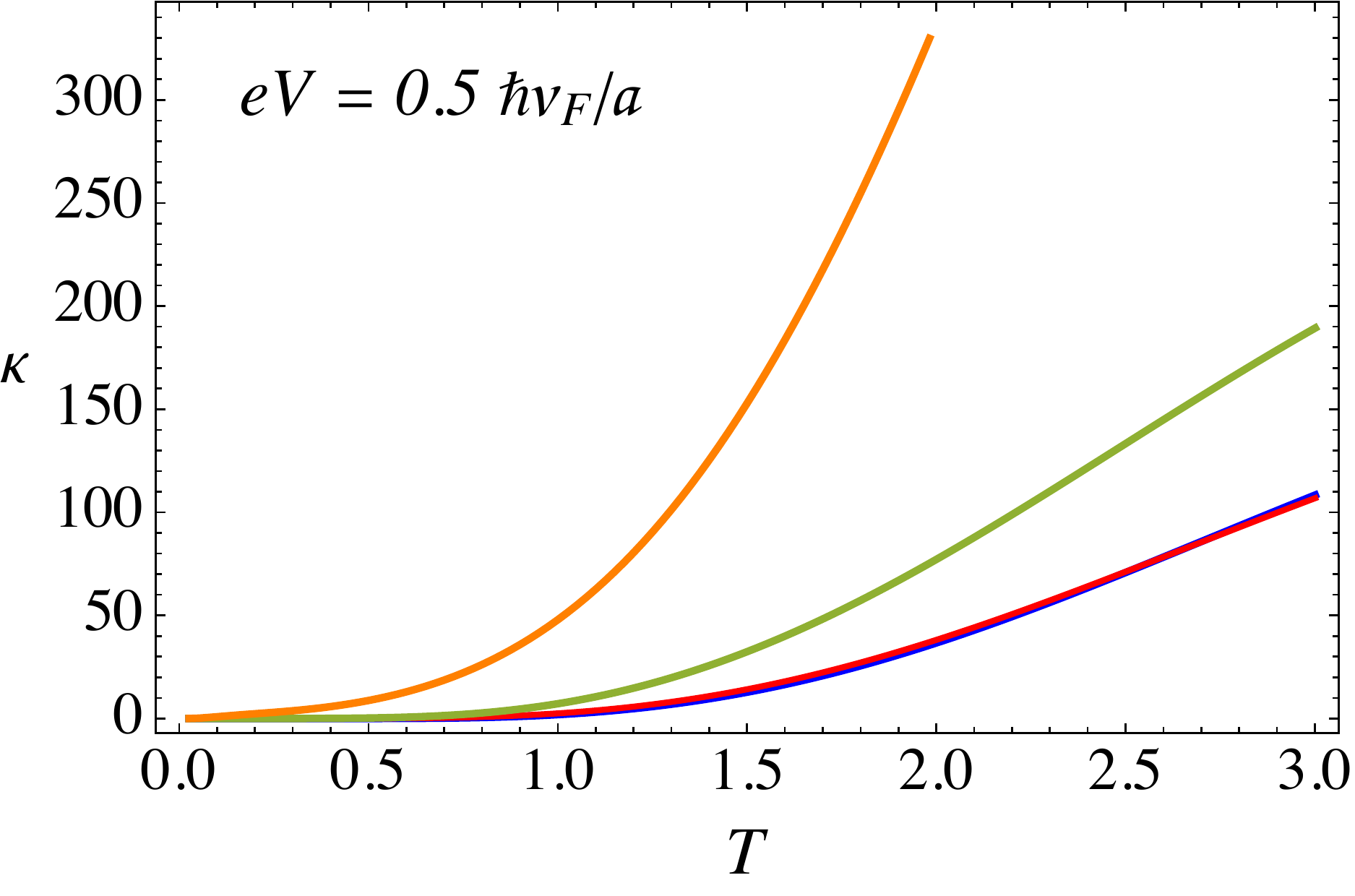}\label{fig:thermal_cond_a}}
	\subfigure[]{\includegraphics[width=0.48\textwidth]{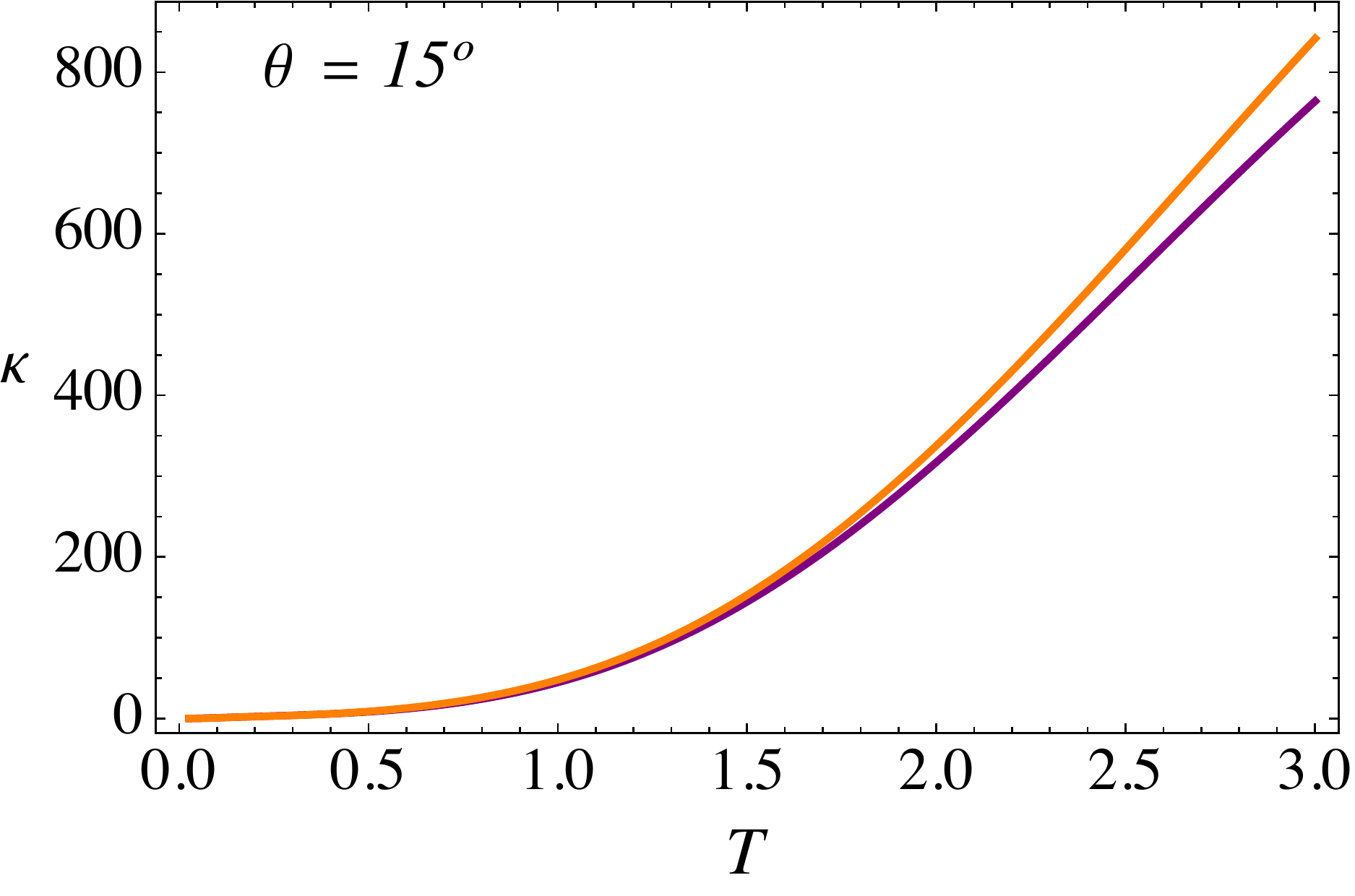}\label{fig:thermal_cond_b}}
	\caption{(Color online)(a) Thermal conductance (in units of $k_Bv_F/a$) as a function of temperature (in units of $\hbar v_F/k_B a$), computed from the analytical expression in Eq.\eqref{eq:kappa_final_mag_torsion_RDSP}, for external $B_0a^2=25\tilde{\phi}_0$, a bias $e V = 0.5\, \hbar v_F/a$ and $\alpha=3\pi/4$. The blue line corresponds to $\theta=0^{\circ}$, red is for $\theta=5^{\circ}$, green is for $\theta=10^{\circ}$ and the orange line corresponds to $\theta=15^{\circ}$. (b) Comparison of the thermal conductance for $\theta=15^{\circ}$: the purple line is for $\alpha=0$ whereas the orange line is for $\alpha=3\pi/4$.}
	\label{fig:thermal_cond}
\end{figure}

Let us now analyze the thermoelectric transport coefficients. Fig.~\ref{fig:conductanceT_a} presents the electric conductance (in units of $e^2/\hbar$) as a function of temperature (in units of $\hbar v_F/k_B a$) for an external field $B_0a^2=25\tilde{\phi}_0$, a bias voltage $e V = 0.5\, \hbar v_F/a$, $\alpha=3\pi/4$, and different torsion angles $\theta$. On the other hand, Fig.~\ref{fig:thermal_cond_a} presents the thermal conductance (in units of $e^2/\hbar$) as a function of temperature, for the same set of parameters as in Fig.~\ref{fig:conductanceT_a}. Both transport coefficients show a monotonic increase with temperature. This effect occurs because Weyl fermions are the same entities transporting current and energy, since as we explained in Sec.~Theory, in the present work we only consider the electronic contribution to the transport. Other effects, such as phonons, will be analyzed in future work. 

From Fig.~\ref{fig:conductanceT_a} and Fig.~\ref{fig:thermal_cond_a}, it is clear that both transport coefficients, i.e. the thermal and the electric conductance, increase with torsion. This effect, already observed in our previous work in the absence of the lattice mismatch barrier contribution\cite{Munoz2019}, is due to the enhancement of the pseudo-Landau levels density of states arising from the $\xi = -1$ chiral node, as already discussed in the previous section.  \\

\begin{figure}
	\centering
	\subfigure[]{\includegraphics[width=0.48\textwidth]{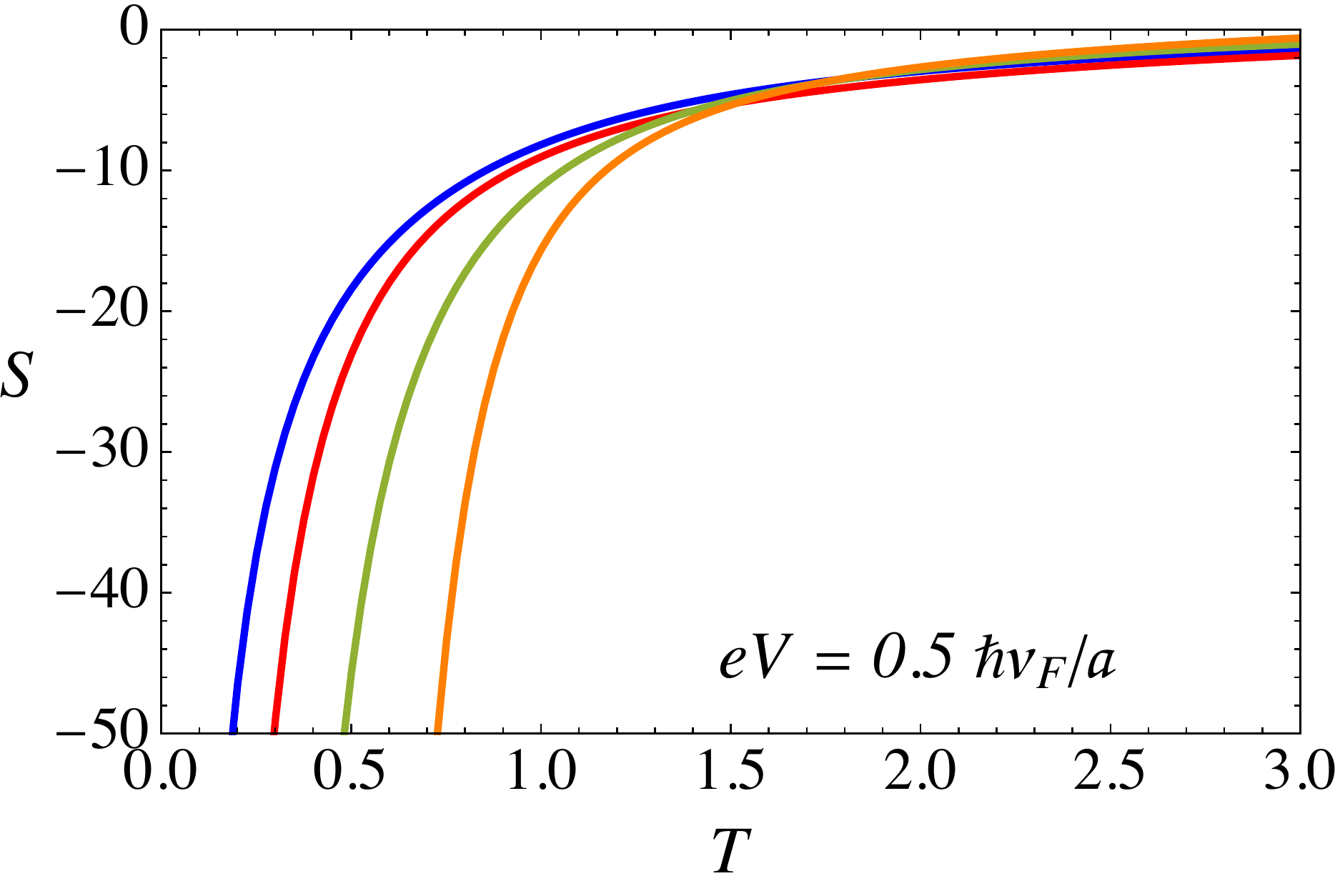}\label{fig:seebeck_a} }
	\hskip .1cm 
	\subfigure[]{\includegraphics[width=0.48\textwidth]{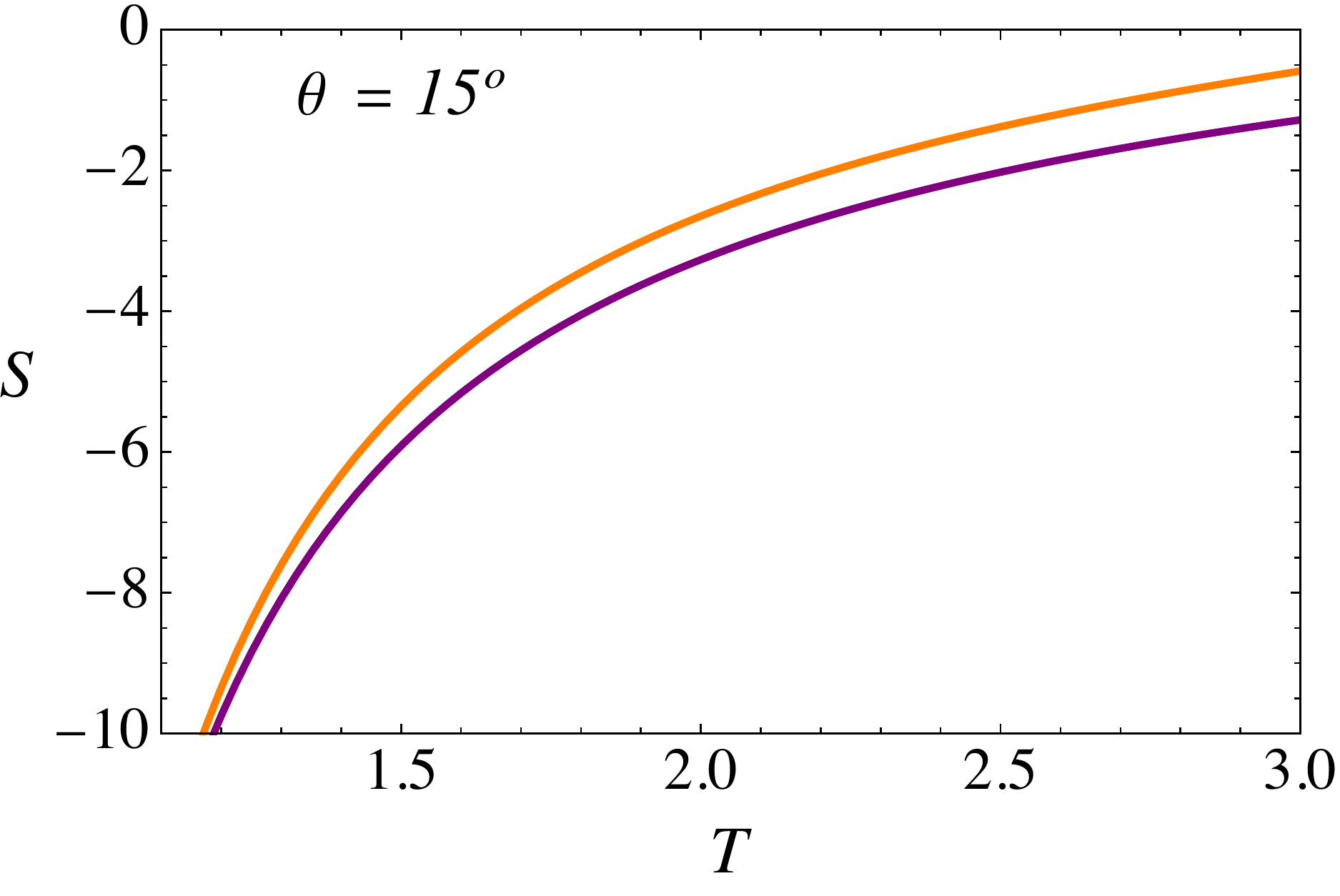}\label{fig:seebeck_b}}
	\caption{(Color online)(a) Seebeck coefficient (in units of $k_B/e$)  computed from the analytical expression in Eq.\eqref{eq:Seebeck_mag_torsion_RDSP} as a function of temperature $T$ (in units of $\hbar v_F/k_B a$). The plot is for fixed $B_0a^2=25\tilde{\phi}_0$, a bias $e V = 0.5\, \hbar v_F/a$ and $\alpha=3\pi/4$. The blue line corresponds to $\theta=0^{\circ}$, red is for $\theta=5^{\circ}$, green is for $\theta=10^{\circ}$ and the orange line corresponds to $\theta=15^{\circ}$. (b) Comparison of the Seebeck coefficient for $\theta=15^{\circ}$: the purple line is for $\alpha=0$ whereas the orange line is for $\alpha=3\pi/4$.}
	\label{fig:seebeck}
\end{figure}

Fig.~\ref{fig:seebeck_a} shows the Seebeck coefficient (in units of $k_B/e$) as a function of temperature (in units of $\hbar v_F/k_B a$),  for the same set of parameters as in Figs.~\ref{fig:conductanceT_a} and \ref{fig:thermal_cond_a}. We have chosen the chemical potential as $\mu=1.0\,\hbar v_{F}/a>0$, such that the negative charge carriers dominate the transport, which explains the negative sign of the Seebeck coefficient. As can be seen, the slope of $S$ is very steep at low temperatures and varies monotonically. On the other hand, the absolute value of $S$ increases with the torsion angle $\theta$.

Now, let us discuss the effect of the RDSP barrier representing the lattice mismatch via the parameter $\alpha$. Figs.~\ref{fig:conductanceT_b} and \ref{fig:thermal_cond_b} present a comparison of the $\alpha=0$ and $\alpha=3\pi/4$ cases for the electric and thermal conductance, respectively, as a function of temperature. For the case of electric conductance the effect is hardly noticeable, with a tiny decrease of the conductance for the case with the RDSP barrier present, $\alpha \ne 0$. On the contrary, the effect is most notorious for the case of the thermal conductance, which increases when the lattice mismatch barrier is present. In both cases, the effect tends to be more significant at high temperatures. \\

\begin{figure}[H]
	\centering
	\subfigure[]{\includegraphics[width=0.48\textwidth]{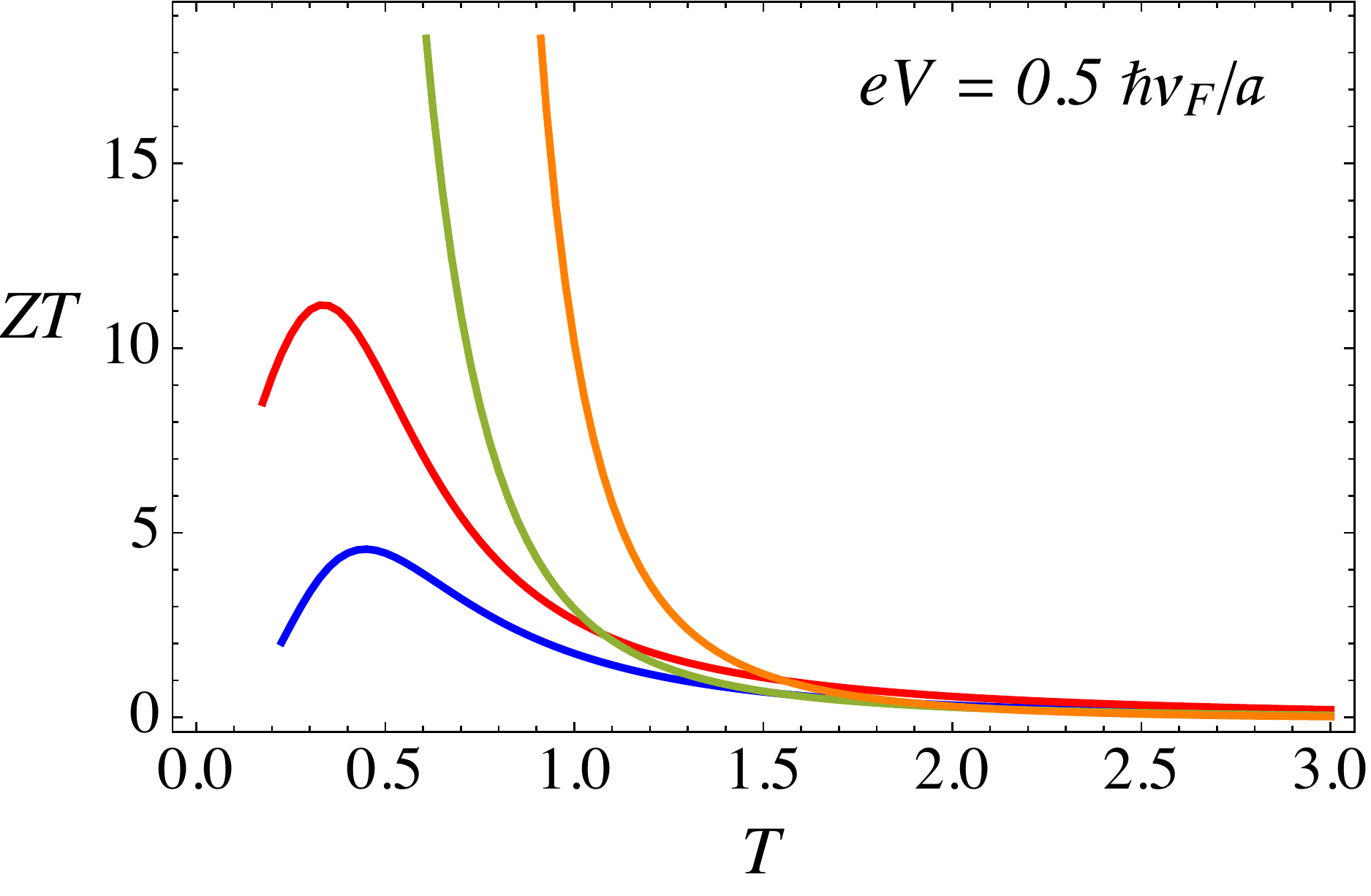}\label{fig:meritZT_a} }
	\hskip .1cm 
	\subfigure[]{\includegraphics[width=0.48\textwidth]{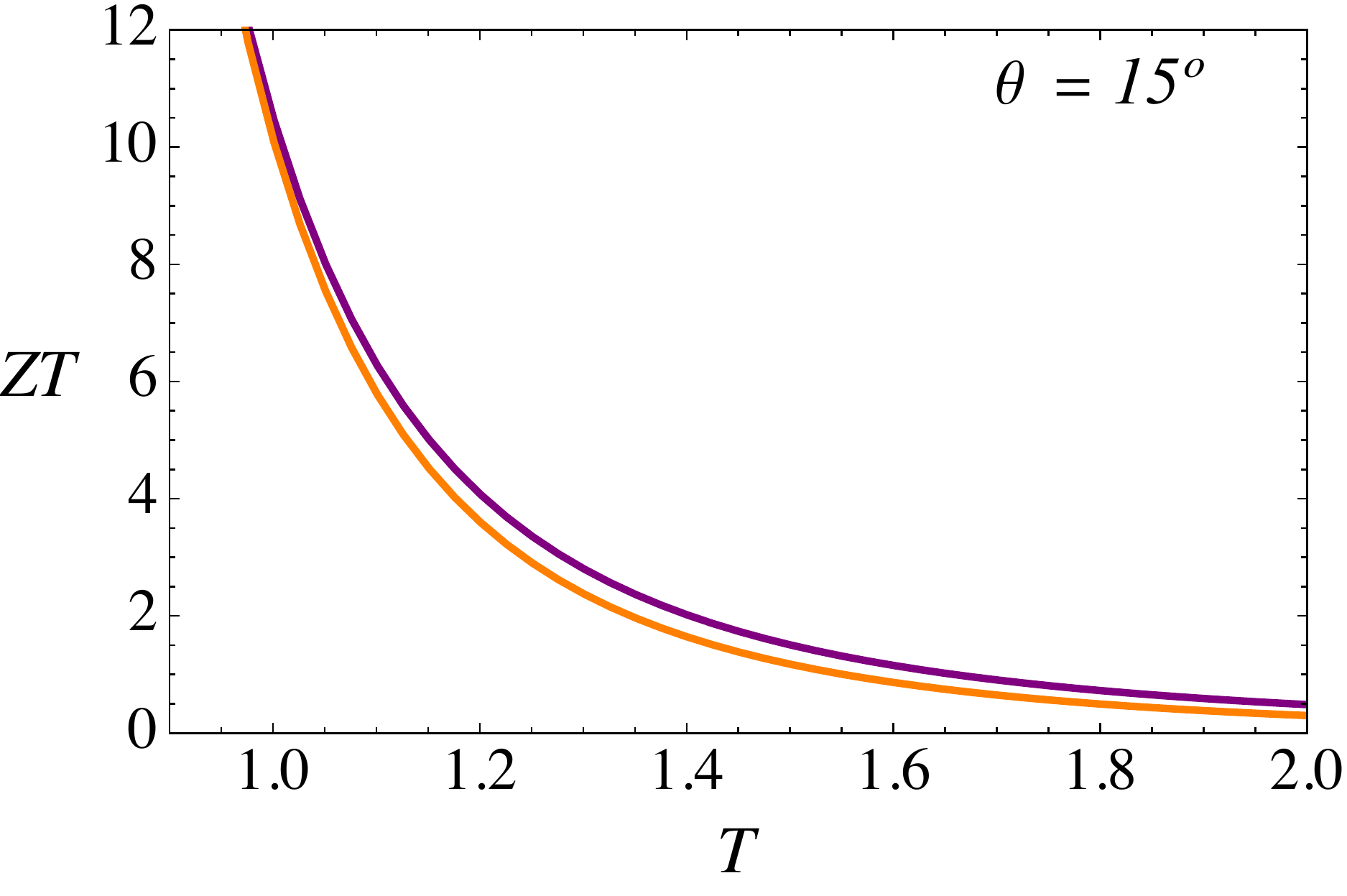}\label{fig:meritZT_b}}
	\caption{(Color online)(a) The figure of merit ZT (dimensionless) as a function of temperature (in units of $ \hbar\, v_F/k_Ba$), calculated for fixed $B_0a^2=25\tilde{\phi}_0$, a bias $e V = 0.5\, \hbar v_F/a$ and $\alpha=3\pi/4$. The blue line corresponds to $\theta=0^{\circ}$, red is for $\theta=5^{\circ}$, green is for $\theta=10^{\circ}$ and the orange line corresponds to $\theta=15^{\circ}$. (b) Comparison of the figure of merit ZT for $\theta=15^{\circ}$: the purple line is for $\alpha=0$ whereas the orange line is for $\alpha=3\pi/4$.}
	\label{fig:meritZT}
\end{figure}

For the characterization of the thermoelectric performance of this WSM junction, a useful quantity is the magnitude of the figure of merit $ZT$, defined by the well known formula 
 \begin{equation}
 ZT = S^2  \frac{T\,G(T,V)}{\kappa(T,V)}. \label{eq:meritZT}
 \end{equation}
Fig.~\ref{fig:meritZT_a} presents the figure of merit $ZT$ (dimensionless), as a function of temperature and for various torsion angles $\theta$. As we showed in our previous work in the absence of the lattice mismatch effect~\cite{Munoz2019}, it is important to notice that extremely high values of $ZT$ can be achieved through the combination of external magnetic field and torsional strain. The value of $ZT$ increases with the torsion angle $\theta$, and the effect is more appreciable at low temperatures. The effect of the RDSP barrier representing lattice mismatch by the parameter $\alpha$ is shown in Fig.~\ref{fig:meritZT_b}. The presence of the barrier produces a small reduction of the figure of merit at high temperatures. 

\begin{figure}[H]
	\centering
	\subfigure[]{\includegraphics[width=0.48\textwidth]{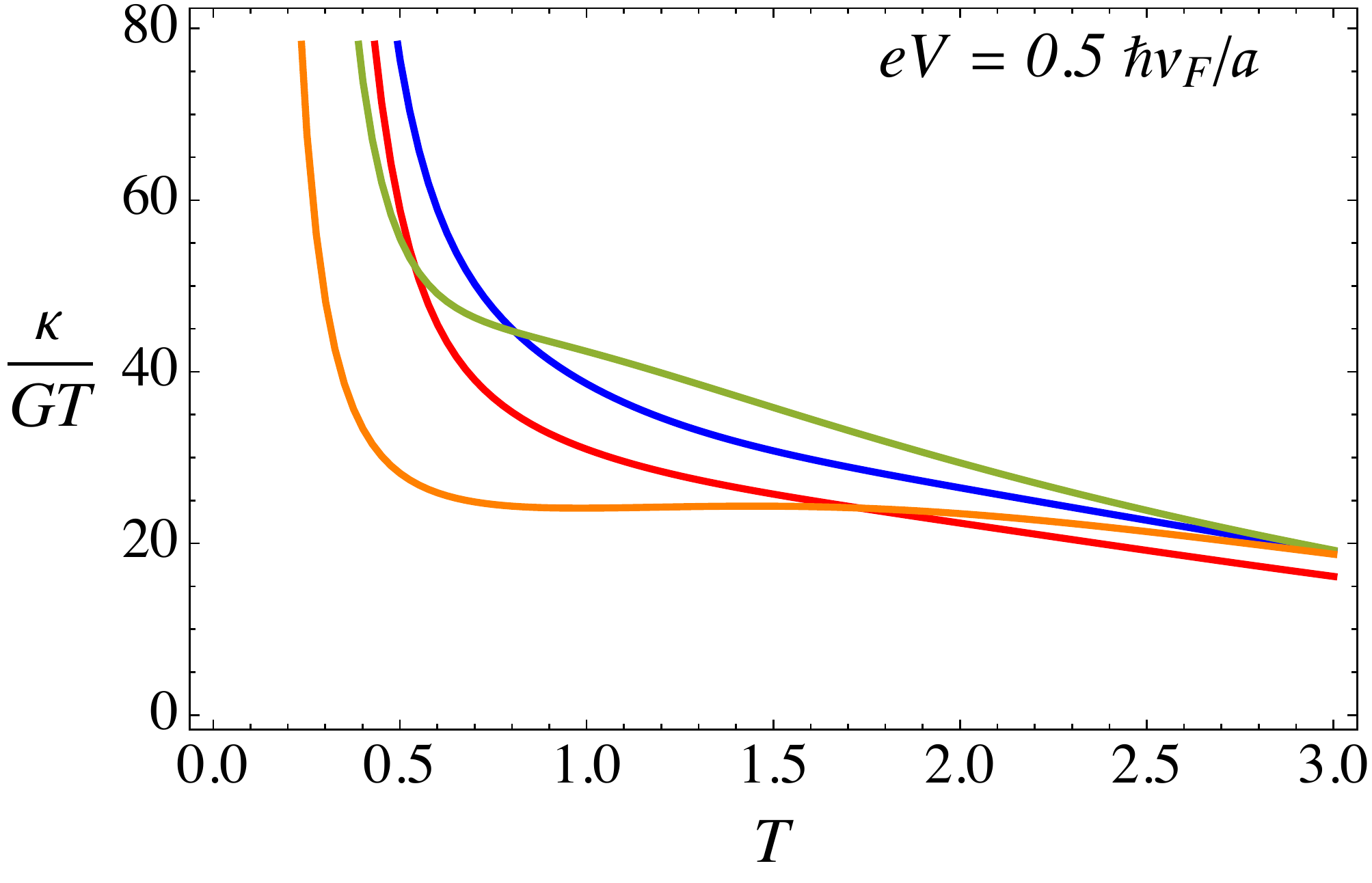}\label{fig:lorentz_a}}
	\hskip .1cm 
	\subfigure[]{\includegraphics[width=0.48\textwidth]{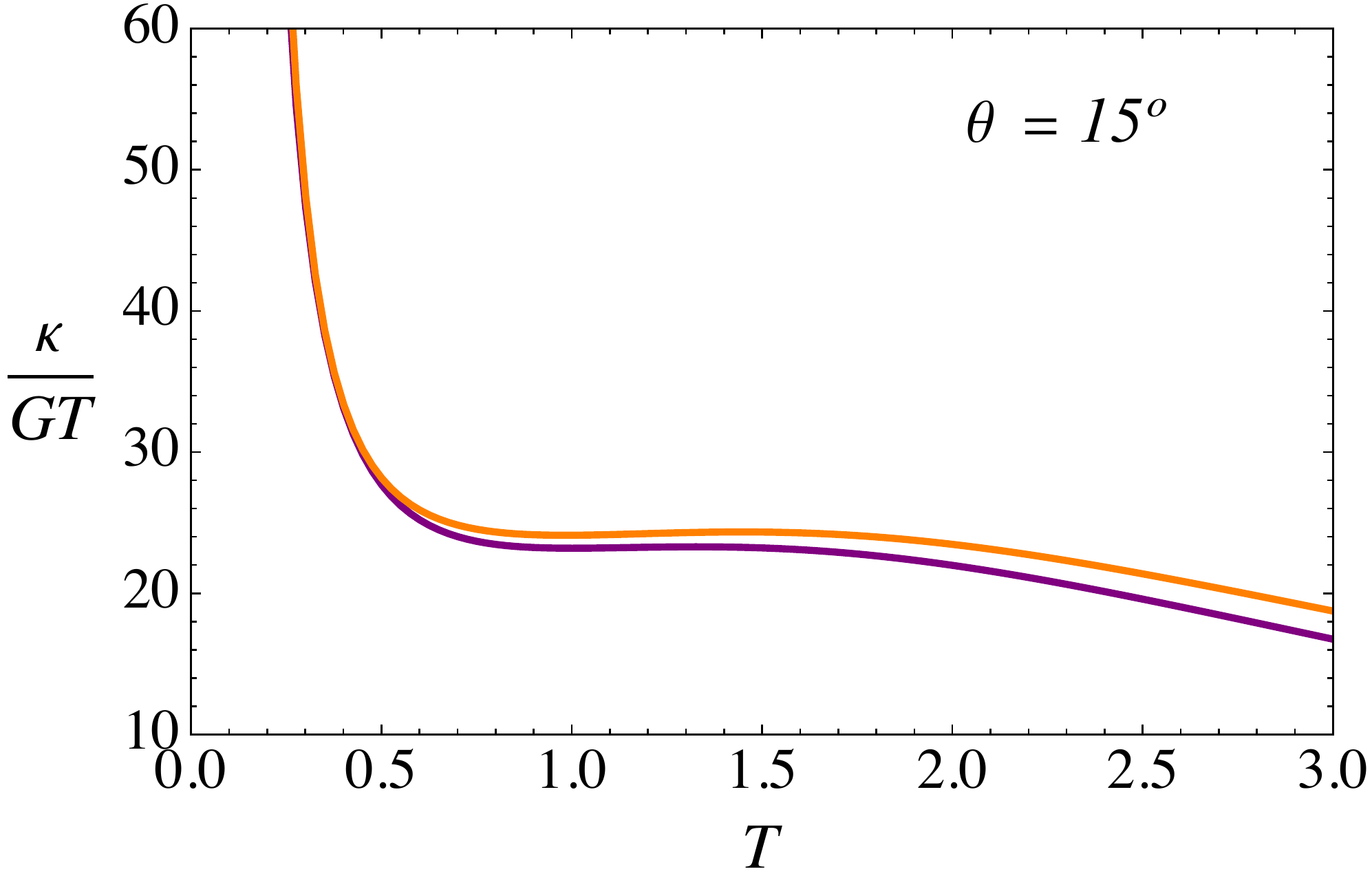}\label{fig:lorentz_b}}
		\caption{(Color online) (a) The Lorenz number (in units of $k_B^2/e^2$) as a function of temperature (in units of $\hbar\, v_F/k_Ba$), calculated for fixed $B_0a^2=25\tilde{\phi}_0$, a bias $e V = 0.5\, \hbar v_F/a$ and $\alpha=3\pi/4$. The blue line corresponds to $\theta=0^{\circ}$, red is for $\theta=5^{\circ}$, green is for $\theta=10^{\circ}$ and the orange line corresponds to $\theta=15^{\circ}$. (b) Comparison of the Lorenz number for $\theta=15^{\circ}$: the purple line is for $\alpha=0$ whereas the orange line is for $\alpha=3\pi/4$.}
	\label{fig:lorentz}
\end{figure}

It is also pertinent to explore the deviation from the metallic behavior by studying the Lorenz number as a function of temperature. The Lorenz number is defined by the formula
 \begin{equation}
 L = \frac{\kappa(T,V)}{T G(T,V)}.\label{eq:Lorenz}
 \end{equation}
 The Lorenz number is represented, at fixed bias and magnetic field, as a function of temperature for different values of torsion in Fig.~\ref{fig:lorentz_a}. Strong deviations from the Wiedemann-Franz law are observed at low temperatures. This effect occurs because the electronic conductance exhibits a non-metallic behavior at low temperatures, due to the discrete pseudo-Landau level spectrum, as can be seen in the staircase pattern in Fig.~\ref{fig:currentsT0}.  It is precisely this effect that explains the extremely high
 $ZT$ values at low temperatures, in agreement with the experimental evidence reported \cite{Skinner} that suggested values as high as $ZT \sim 10$. In contrast with the $ZT$ behavior, the presence of the delta barrier is to increase the Lorenz number at high temperatures, as can be seen in Fig.~\ref{fig:lorentz_b}. This trend is explained since, as we discussed previously, at high temperatures the thermal conductance increases with the delta barrier, while the electric conductance slightly decreases.  

%%%%%%%%%%%%%%%%%%%%%%%%%%%%%%%%%%%%%%%%%%
\section{Discussion}

In this work, we studied the thermoelectric transport properties of a type I Weyl semimetal with a torsional defect, in the presence of an external magnetic field along the axis of the dislocation in a cylindrical geometry. Moreover, the effect of torsion was modeled by a combination of a gauge field representation, and a repulsive delta-shell potential (RDSP) representing the lattice mismatch at the edge of the cylindrical region. We remark that the mechanical gauge field, in combination with the external magnetic field imposed upon the region, combine into an effective node-dependent pseudo-magnetic field $\mathbf{B}^{\xi} = \mathbf{B} + \xi \mathbf{B}_S$ (for $\xi = \pm)$ that breaks time-reversal symmetry and hence the nodal symmetry. Therefore, our analysis shows that the electronic states within the region correspond to effective node-polarized Landau levels, leading to a node-polarization effect of the total electric current $I = I_{+} + I_{-}$. In particular, the low-temperature differential conductance displays the corresponding characteristic trend of discrete peaks corresponding to each of such Landau levels. We also demonstrated that the effect of the lattice-mismatch, represented by the RDSP, is periodic in the strength of the repulsive barrier $V_0$, in the form $\tan(V_0/\hbar v_F)$, thus revealing the presence of "magic angles" (the zeroes of the tangent) where the barrier becomes transparent. This somewhat surprising effect is a manifestation of the Klein-tunneling effect of Dirac's theory, observed in this particular context and geometry. Finally, we also studied the thermoelectric transport coefficients, thermal conductivity and Seebeck, as a function of temperature, external magnetic field, torsion and strength of the lattice mismatch (RDSP). 

{\color{black} We would like to emphasize that our analytical equations, and the corresponding figures presented in the Results section, are expressed in terms of dimensionless groups involving structural parameters (such as the radius $a$ of the torsional defect and the dimensions $W$ and $L$ of the WSM slab) as well as the material's specific parameters (such as the Fermi velocity $v_F$). This has the advantage that the equations presented are quite general, and hence our theoretical predictions for the transport coefficients can be compared with specific experimental measurements by choosing the appropriate material-dependent parameters. For instance, choosing the dimensions of the slab as $W\sim L \sim 50 \,\,\text{nm}$ and the radius of the cylindrical strip as $a\sim 15\,\, \text{nm}$, we obtain an electrical resistivity $\rho\sim 2.15\times 10^{-4}\,\,\Omega\text{m}$ which is within the range reported in Ref. \cite{han2020} ($\rho\sim 2\times 10^{-2}\,\,\Omega\text{m}$ for Bi and $\rho\sim  10^{-5}\,\,\Omega\text{m}$ for TaP). On the other hand, for the case of the thermal conductivity, using the Fermi velocity $v_F\sim 1.5 \times 10^6 \,\,\text{m}/\text{s}$ for the material Cd$_{3}$As$_2$ \cite{Neupane2014}, and the same values for $a$, $L$, and $W$ as before we found a value of $\kappa \sim 6.6\,\,\text{W}/\text{mK}$ which is of the same order of magnitude to those reported in Ref. \cite{Skinner} ($\sim 3 \,\,\text{W}/\text{mK}$ for Pb$_{1-x}$Sn$_x$Se) and in Ref. \cite{han2020} ($\sim 5-25 \,\,\text{W}/\text{mK}$ for TaP)}.

Finally, we point out that our theoretical calculations suggest that a very high figure of merit can be obtained from such configuration (torsional strain + RDSP), thus constituting a very interesting candidate for thermoelectric applications in energy harvesting.

%%%%%%%%%%%%%%%%%%%%%%%%%%%%%%%%%%%%%%%%%%
%\section{Conclusions}

%This section is not mandatory, but can be added to the manuscript if the discussion is unusually long or complex.

%%%%%%%%%%%%%%%%%%%%%%%%%%%%%%%%%%%%%%%%%%
\vspace{6pt} 

%%%%%%%%%%%%%%%%%%%%%%%%%%%%%%%%%%%%%%%%%%
%% optional
%\supplementary{The following are available online at \linksupplementary{s1}, Figure S1: title, Table S1: title, Video S1: title.}


%%%%%%%%%%%%%%%%%%%%%%%%%%%%%%%%%%%%%%%%%%
% \authorcontributions{E. M. and R. S.-G. conceptualized the idea; D. B. and E. M. performed the analytical calculations and wrote the manuscript; D. B. performed the numerical calculations; R. S.-G. reviewed and edited the manuscript; All authors have read and agreed to the published version of the manuscript.}

\funding{This research was funded by Fondecyt grants number 1190361 and 1200399, as well as by ANID PIA Anillo ACT/192023.}

%%%%%%%%%%%%%%%%%%%%%%%%%%%%%%%%%%%%%%%%%%

\conflictsofinterest{The authors declare no conflict of interest.}

%%%%%%%%%%%%%%%%%%%%%%%%%%%%%%%%%%%%%%%%%%
%% Only for journal Encyclopedia
%\entrylink{The Link to this entry published on the encyclopedia platform.}

%%%%%%%%%%%%%%%%%%%%%%%%%%%%%%%%%%%%%%%%%%
%% Optional
\abbreviations{The following abbreviations are used in this manuscript:\\

\noindent 
\begin{tabular}{@{}ll}
WSM & Weyl semimetal\\
RDSP & Repulsive delta-shell potential\\
\end{tabular}}

%%%%%%%%%%%%%%%%%%%%%%%%%%%%%%%%%%%%%%%%%%

%%%%%%%%%%%%%%%%%%%%%%%%%%%%%%%%%%%%%%%%%%
\end{paracol}
\reftitle{References}

\end{document}